\documentclass{aa}

\usepackage{graphicx}
\usepackage[normalem]{ulem}
\usepackage{txfonts}
\usepackage{tablefootnote}
\usepackage{threeparttable}
\usepackage{booktabs}
\usepackage{bm}
\usepackage{graphicx}
\usepackage{mathtools}
\usepackage{natbib}
\usepackage{longtable}
\usepackage[colorinlistoftodos]{todonotes}
\usepackage{subcaption}
\usepackage[T1]{fontenc}
\usepackage{lscape}
\newcommand{\bd}{BD+46$^{\circ}$442}

\newcommand{\iras}{IRAS19135+3937}

\newcommand{\mdot}{M_\odot}
\newcommand{\myr}{M_\odot\,\text{yr}^{-1}}

\newcommand{\kms}{km\,s$^{-1}$}
\newcommand{\halpha}{H$\alpha$}
\newcommand{\hbeta}{H$\beta$}
\newcommand{\hgamma}{H$\gamma$}
\newcommand{\hdelta}{H$\delta$}

\begin{document}

   \title{The structure of jets launched from post-AGB binary systems\thanks{Based on observations made with the Mercator Telescope, operated on the island of La Palma by the Flemish Community at the Spanish Observatorio del Roque de los Muchachos of the Instituto de Astrofísica de Canarias.}}

     \author{D. Bollen \inst{1,2,3}, D. Kamath \inst{2,3}, H. Van Winckel \inst{1}, and O. De Marco \inst{2,3}, O. Verhamme \inst{1}, J. Kluska \inst{1}, and M. Wardle \inst{2,3}
       }
    
            \institute{
                  Institute of Astronomy, KU Leuven, Celestijnenlaan 200D, 
 B-3001 Leuven, Belgium
                   \email{hans.vanwinckel@kuleuven.be}
                  \and
                  Department of Physics \& Astronomy, School of Mathematical and Physical Sciences, Macquarie University, 
                  Sydney, NSW 2109, Australia
                  \and
                  Astronomy, Astrophysics and Astrophotonics Research Centre, 
                  Macquarie University, Sydney, NSW 2109, Australia
                  }
    
       \date{}
       \authorrunning{Bollen et al.}
 
  \abstract
   {In this paper, we focus on post-asymptotic giant branch (post-AGB) binaries and study the interaction between the different components of these complex systems. These components comprise the post-AGB primary, a main sequence secondary, a circumbinary disk, as well as a fast bipolar outflow (jet) launched by the companion. We obtained well-sampled time series of high resolution optical spectra over the last decade and these spectra provide the basis of our study. }
   { We aim to use the time-series data to quantify the velocity and density structure of the jets in nine of these post-AGB binaries. This complements our earlier work and this amounts to the analyses of 16 jet-launching systems in total. }
   {The jet is detected in absorption, at superior conjunction, when the line of sight towards the primary goes through the bipolar cone. Our spectral time series scan the jets during orbital motion. Our spatio-kinematic model is constrained by these dynamical spectra.  We complement this with a radiative-transfer model in which the Balmer series are used to derive total mass-loss rates in the jets. }
   {The jets are found to be wide ($>30\degr$)  and display an angle-dependent density structure with a dense and slower outer region near the jet cone and a fast inner part along the jet symmetry axes. The deprojected outflow velocities confirm that the companions are main sequence companions. The total mass-loss rates are large ($10^{-8}-10^{-5}\,\myr$), from which we can infer that the mass-accretion rates onto the companion star must be high as well. The circumbinary disk is likely the main source for the accretion disk around the companion. 
   All systems with full disks that start near the sublimation radius show jets, whereas for systems with evolved transition disks this lowers to a detection rate of 50\%. Objects without an infrared excess do not show jets.
    }
   {We conclude that jet creation in post-AGB binaries is a mainstream process. Our geometric spatio-kinematic model is versatile enough to model the variety of spectral time series. The interaction between the circumbinary disks and the central binary provide the needed accretion flow, but the presence of a circumbinary disk does not seem to be the only prerequisite to launch a jet.   }

   \keywords{Stars: AGB and post AGB     
           Stars: binaries: spectroscopic      
           Stars: circumstellar matter   -
           Stars: mass loss --
           ISM: jets and outflows --
           Accretion, accretion disks
           }

   \maketitle
%

\section{Introduction}\label{sec:introduction}

    Astrophysical collimated mass outflows or jets cover a wide range of energies, from ultra-relativistic jets powered by supermassive black holes in the centres of active galactic nuclei, to stellar jets powered by accretion disks around young stellar objects \citep[YSOs, e.g.][and references therein]{romero2021}. Jets are launched from accreting objects and the jet acceleration and collimation is controlled by a magnetic field \citep[e.g.][]{livio2002}. Modelling jets is challenging because of their non-linear nature, the poorly understood disk–jet connection near their base and turbulent interactions with the ambient medium as they propagate away from the centre. The mechanisms related to jet launching, acceleration and collimation are still heavily debated.
    
\begin{figure}
    	\centering
    	\includegraphics[width=.5\textwidth]{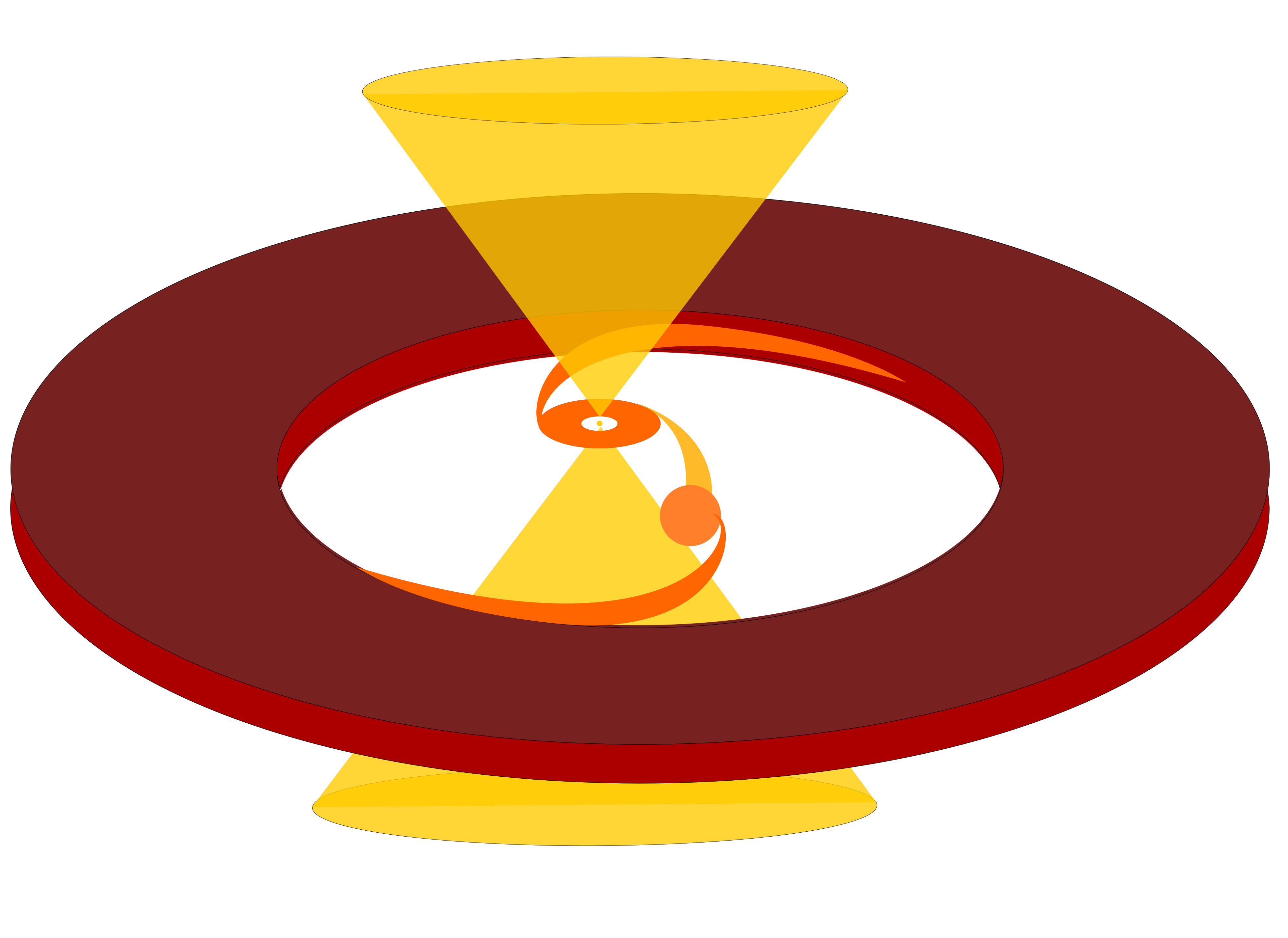}
    	\caption{Cartoon of the main structural elements of post-AGB binaries: the circumbinary disk, the central binary with the primary post-AGB star, and the main sequence companion with its accretion disk which launches the jet. The interactions between these components is also indicated. \label{fig:structure}}
    \end{figure}

    In this paper we focus on jets as detected in evolved low- to intermediate-mass stars in which one of the components is in its post-asymptotic giant branch (post-AGB) evolutionary phase and the secondary is a main sequence star. These binary post-AGB stars have distinct observational properties which were recently reviewed by \cite{vanwinckel17,vanwinckel2018}. 
    In summary, the systems have orbital periods between approximately a hundred and a few
thousand days. The systems are typically surrounded by a circumbinary disk of dust and gas.  A complete catalogue of Galactic disk objects has been published by \cite{kluska22}, which also contains the full description of the spectral energy distributions (SEDs). The main structural components of these systems are depicted in Fig~\ref{fig:structure}.
    
\begin{figure}
    	\centering
    	\includegraphics[width=.5\textwidth]{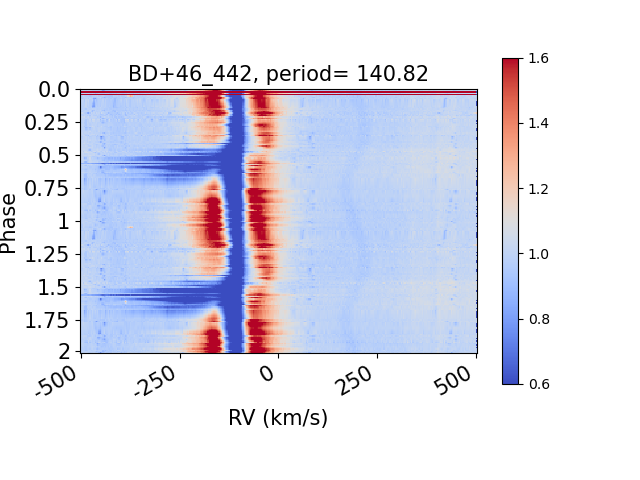}
    	\caption{Dynamic spectra of  \halpha\, in BD+46$^{\circ}$442, folded on the orbital period $-$ two orbital periods are displayed. The spectra are normalised to the continuum. The blueshifted absorption occurs centred around superior conjunction. }
    	\label{fig:example}
    \end{figure}

    The presence of a jet in these post-AGB binary systems is revealed by time series of high resolution optical spectra around the \halpha\, region. In the \halpha\, profile, an extended blue-shifted absorption feature occurs during superior conjunction\footnote{Superior conjunction indicates the phase when the post-AGB star is located behind the companion, as viewed by the observer.} \citep{gorlova12,bollen17}. We illustrate that in Fig.~\ref{fig:example} in the form of a dynamic plot in which we represent our time series for BD+46$^{\circ}$442\, folded on the orbital period. The absorption is seen when the line of sight towards the primary passes through the bicone of the jet, which is launched by the accretion disk around the companion.
    
    Owing to the orbital motion of the binary, we scan through the jet during orbital motion and hence probe the structure and velocity of the material within the jet. The line of sight towards the primary and the orbital geometry determines which part of the jet is scanned. These systems provide a rare opportunity in which time-resolved spectra during orbital motion can be used to investigate the jet dynamics and jet structure.
    
    This paper is embedded in a series in which we model in detail the spectral time series of a number of post-AGB binaries with jets, to investigate the jet structures. We introduce the sample of objects and their time series in Sect~\ref{sec:sample} and explain our modelling algorithm in Sect~\ref{sec:methods}. We refer the reader to our earlier modelling of a smaller sample of stars in \cite{bollen20} as Paper~I  and \cite{bollen21} as Paper~II. We highlight our most important results on this extensive sample study of 9 objects in Sect~\ref{sec:results}. Next, we  draw conclusions on the jet structure, the mass-loss rates, and the accretion rates from the full sample of 16 objects analysed so far. Finally, we evaluate the results in the context of the whole sample of Galactic post-AGB binaries and formulate the next steps in our research.


\section{Sample selection and observational data}\label{sec:sample}
    
    The long-term monitoring campaign with the HERMES spectrograph \citep{raskin11}, mounted on the Mercator telescope started in 2009 and has resulted in a large collection of high-resolution optical spectra of post-AGB systems \citep{vanwinckel09}. Our observations are drawn from this campaign.
    
    Here, we present a detailed description of our global target selection. We began with all the post-AGB binaries for which the orbital period is determined. A recent study by \cite{oomen18} performed a radial velocity analysis of 33 Galactic post-AGB binary systems, with the aim of updating their orbital parameters. We collected spectra for these 33 systems, as well as from three additional objects: RV~Tau \citep{manick19}, SU~Gem, and the W\,Vir star AU Peg \citep{gonzalez96}. Seven of these objects have already been modelled in Paper~I (\bd, \iras) and Paper~II (89~Her, EP~Lyr, HD~46703, HP~Lyr, and TW~Cam). For several of these object (89~Her, HD~46703, and TW~Cam), we obtained additional observations since those publications and we have here updated the jet parameters by re-modelling those objects. The Mercator telescope is located at the Roque de los Muchachos observatory (latitude $+28^{\circ}45'24"\,$ N). Thus, we further focus on the northern objects that are much better sampled than the southern ones.
    
    Next, we narrow down our selection based on two main criteria: the presence of a jet absorption feature in the spectra and a good coverage of the orbital phase.

The blue-shifted \halpha\, absorption appears during superior conjunction in almost all objects of the sample of \cite{oomen18}. Only five objects do not show a clear jet absorption feature (BD+39${^\circ}$4926, DY\,Ori, AU\,Peg, IRAS05208$-$2035, HD~137569). We also omitted HD\,44179, which is the central star of the Red Rectangle. We see the Red Rectangle almost edge on \citep[e.g.][]{cohen04}. We also refer to \cite{thomas13} for a study of the dynamics of the inner nebula including the presence of a jet. As such, we removed these objects from our sample. We include the post-AGB binaries RV~Tau and SU~Gem, which were not part of the study by \cite{oomen18} as they also have a significant absorption feature.
    	
We aim for a minimum of 20 spectra per object, in order to ensure a good coverage of the orbital phase and the jet absorption feature. If the coverage of the orbital phase is poor, the model could over fit the observed phases. We limit ourselves to the objects observable by the Mercator telescope. Additionally, since we are interested in the jet absorption feature, we require good sampling of the orbital phases where the jet absorption is present. Since the absorption often displays cycle-to-cycle variability, we choose not to use all the available spectra. Instead, we select between 20 and 60 spectra per object. We found that this number of spectra is optimal to be unaffected by cycle-to-cycle variability and to reduce computing time, whilst still being constraining enough for the modelling.

    Our selection criteria result in a final sample of nine jet-creating post-AGB binary systems with good orbital phase coverage and for which we did not publish results yet. We list our programme stars and their orbital elements in Table~\ref{tab:tableorbitalelements}. Table~\ref{tab:tableorbitalelements2} includes also information of the full sample of 16 objects. The orbital information is given as well as the effective temperature and luminosity of the post-AGB star, the luminosity determined from the GAIA distances, the infrared luminosity, the depletion pattern of the post-AGB star, the total number of spectra that are used in this analysis, and their average signal-to-noise ratio.

\begin{table*}[ht]
	\centering
	\caption{Orbital elements of the binary systems.}
	\label{tab:tableorbitalelements}
	\begin{tabular}{llccccccc} 
		\hline \hline
		\# & Object           & Period           & Eccentricity     & $T_0$             & $\omega$       & $K_1$          & $\gamma$          & ref.   \\
		&  & days              &                  & days              & Degrees        & km$\,s^{-1}$    & km$\,s^{-1}$      &\\
		\hline
		1 &  AC~Her          & 1188.9$\pm$1.2    & 0.0$+$0.05       &                   &                & 10.8$\pm$0.7   & -28.8$\pm$0.5  &  1\\
		2 &  HD~213985       & 259.6$\pm$0.7     & 0.21$\pm$0.05    & 2407110$\pm$16    & 105$\pm$26   & 31.4$\pm$1.0   & -42.0$\pm$0.9  & 1 \\
		3&  HD~52961        & 1288.6$\pm$0.3    & 0.23$\pm$0.01    & 2407308$\pm$24    & 297$\pm$6  & 13.1$\pm$0.3   & 6.2$\pm$0.2     &   1 \\
		4 &  IRAS~06165+3158 & 262.6$\pm$0.7     & 0.0$+$0.05       &    		       &                & 15.5$\pm$0.5   & -16.4$\pm$0.3   & 1\\
		5 &  IRAS~19125+0343 & 519.7$\pm$0.7     & 0.24$\pm$0.03    & 2451503$\pm$11    & 243$\pm$8  & 12.0$\pm$0.5   & 67.3$\pm$0.3 &  1  \\
		6 &  RV~Tau          & 1198$\pm$17       & 0.5$\pm$0.1      & 2455114$\pm$12    & 4.9$\pm$0.3    & 14.5$\pm$1.6   & 32.0$\pm$0.8 &  2  \\
		7 &  SAO~173329      & 115.951$\pm$0.002  & 0.0$+$0.04      &                   &                & 12.4$\pm$0.3   & 73.3$\pm$0.2    & 1 \\
		8 &  SU~Gem          & 692.2$\pm$1.4     & 0.07$\pm$0.04    & 2455160$\pm$40    & 22$\pm$6       & 20$\pm$3   & 4.4$\pm$0.2  & 3\\
		9 &  U~Mon           & 2550$\pm$143      & 0.25$\pm$0.06    & 2451988$\pm$316   & 87$\pm$15    & 14.9$\pm$1.1   & 24.1$\pm$1.0   & 1 \\
		\hline
	\end{tabular}
	\tablefoot{The orbital data were obtained from (1)~\cite{oomen18}, (2)~\cite{manick19}, and (3)~this work.}	
\end{table*}

\begin{sidewaystable*}[] 
	\centering
	\caption{Projected semi-major axis, mass functions, effective temperature of the post-AGB star, luminosity of the post-AGB star, infrared luminosity as a function of post-AGB luminosity, SED category \citep{kluska22} , depletion, number of observations used in this work, and the average signal to noise in \halpha. The formal error on the effective temperatures is $250\,$K. Below the line are the objects of Paper~I and ~II.}
	\label{tab:tableorbitalelements2}
	\begin{tabular}{llcc|ccccccccc}
		\hline \hline
		\# &  Object           & $a_1\sin i$      & $f(m)$              & $T_\text{eff}$  & L$_\text{model}$  & L$_\text{GAIA}$ & L$_\text{IR}/$L$_\star$ & SED & Depletion & $N_\text{obs}$  & (S/N)$_\text{avg}$  & ref. \\
		&  & AU               & M$_\odot$            &  K    & $L_\odot$ &  $L_\odot$  &  $L_\odot$ & &  &   &  &\\
		\hline 
		1  &  AC~Her          & 1.18$\pm$0.08     & 0.15$\pm$0.03      &  5800 & $3600\pm600$  & $3500^{270}_{240}$      & 0.21 & 2 & Mild     & 33  & $\sim65$  & 1,6\\
		2  &  HD~213985       & 0.73$\pm$0.03     & 0.78$\pm$0.08      &  8250 & $300\pm40$    & $310^{30}_{20}$         & 0.45 & 1 & Strong   & 25  & $\sim70$  & 1,11\\
		3  &  HD~52961        & 1.51$\pm$0.03     & 0.274$\pm$0.019    &  6000 & $1300\pm200$  & $8000^{2400}_{1700}$    & 0.16 & 1 & Strong   & 42  & $\sim100$ & 1,10\\
		4  &  IRAS~06165+3158 & 0.374$\pm$0.011   & 0.10$\pm$0.01      &  4250 & $600\pm140$   & $1800^{400}_{300}$      & 0.08 & 3 & No       & 28  & $\sim30$  & 1,13\\
		5  &  IRAS~19125+0343 & 0.56$\pm$0.02     & 0.086$\pm$0.010    &  7750 & $15600\pm2000$ & $24000^{12000}_{7000}$& 1.33 & 0 & Strong   & 20  & $\sim55$  & 1,14\\
		6  &  RV~Tau          & 1.38$\pm$0.08     & 0.24$\pm$0.02      &  4810 & $4700\pm1000$& $2100^{400}_{300}$      & 0.39 & 1 & Mild     & 31  & $\sim40$  & 2\\
		7  &  SAO~173329      & 0.132$\pm$0.003   & 0.023$\pm$0.001    &  7000 & $1500\pm200$ & $5500^{2900}_{1700}$    & 0.33 & 1 & No       & 23  & $\sim35$  & 1,15\\
		8  &  SU~Gem          & 0.81$\pm$0.04     & 0.15$\pm$0.02      &  5250 & $1200\pm200$ & $410^{310}_{150}$       & 7.92 & 0 & Mild     & 23  & $\sim35$  & 3,13\\
		9  &  U~Mon           & 3.4$\pm$0.3       & 0.79$\pm$0.18      &  5000 & $3600\pm700$ & $6500^{1500}_{1100}$    & 0.23 & 3 & No       & 36  & $\sim40$  & 1,16\\ \hline
		10  &  89~Her          & 0.106$\pm$0.007   & 0.0019$\pm$0.0004  &  6600 & $4200\pm600$  & $14000^{3400}_{2500}$   & 0.42 & 1 &  Mild     & 26  & $\sim125$ & 1,5\\
		11  &  BD+46~442       & 0.3074$\pm$0.0014 & 0.195$\pm$0.003    &  6250 & $1000\pm160$  & $2600^{1800}_{1000}$    & 0.16 & 1 & No       & 52  & $\sim40$  & 1,7\\
		12 &  EP~Lyr          & 1.30$\pm$0.12     & 0.22$\pm$0.06      &  6200 & $7100\pm1100$ & $10000^{4000}_{2500}$   & 0.02 & 3& Moderate & 30  & $\sim45$  & 1,8\\
		13 &  HD~46703        & 0.839$\pm$0.015   & 0.220$\pm$0.012    &  6250 & $8200\pm1300$ & $7800^{3200}_{2100}$    & 0.02 & 4 &Mild     & 34  & $\sim60$  & 1,9\\
		14 &  HP~Lyr          & 1.27$\pm$0.06     & 0.083$\pm$0.007    &  6300 & $7900\pm1300$ & $6400^{2600}_{1700}$    & 0.91 & 1 & Strong   & 40  & $\sim55$  & 1,12\\
		15 &  IRAS~19135+3937 & 0.209$\pm$0.008   & 0.075$\pm$0.008    &  6000 & $600\pm100$  & $1800^{400}_{300}$      & 0.28 & 1 & No       & 22  & $\sim35$  & 1,13\\
		16  &  TW~Cam          & 0.83$\pm$0.04     & 0.17$\pm$0.02      &  4800 & $3200\pm700$ & $1200^{180}_{140}$      & 0.46 & 1 & No       & 42  & $\sim50$  & 1,16\\
		\hline
	\end{tabular}
	\tablefoot{
		The orbital data were obtained from (1) \cite{oomen18}, (2) \cite{manick19}, and (3) this work. The strength of the depletion patterns were determined by (1) \cite{oomen18} and (4) \cite{oomen19}. The effective temperatures were obtained from (5) \cite{kipper11}, (6) \cite{giridhar98}, (7) \cite{gorlova12}, (8) \cite{gonzalez97a}, (9) \cite{hrivnak08}, (10) \cite{waelkens91}, (11) \cite{deruyter06}, (12) 
 \cite{giridhar05}, (13) \cite{rao14}, (14) \cite{maas05},(15) \cite{rao12}, (16) \cite{giridhar00}. The luminosities were calculated from the stellar radii found in this work (L$_\text{model}$ and from the GAIA parallaxes (L$_\text{GAIA}$). The infrared luminosities and SED categories are from \cite{kluska22}.
	}
\end{sidewaystable*}

\begin{figure*}
    	\centering
    	\includegraphics[width=1.\textwidth]{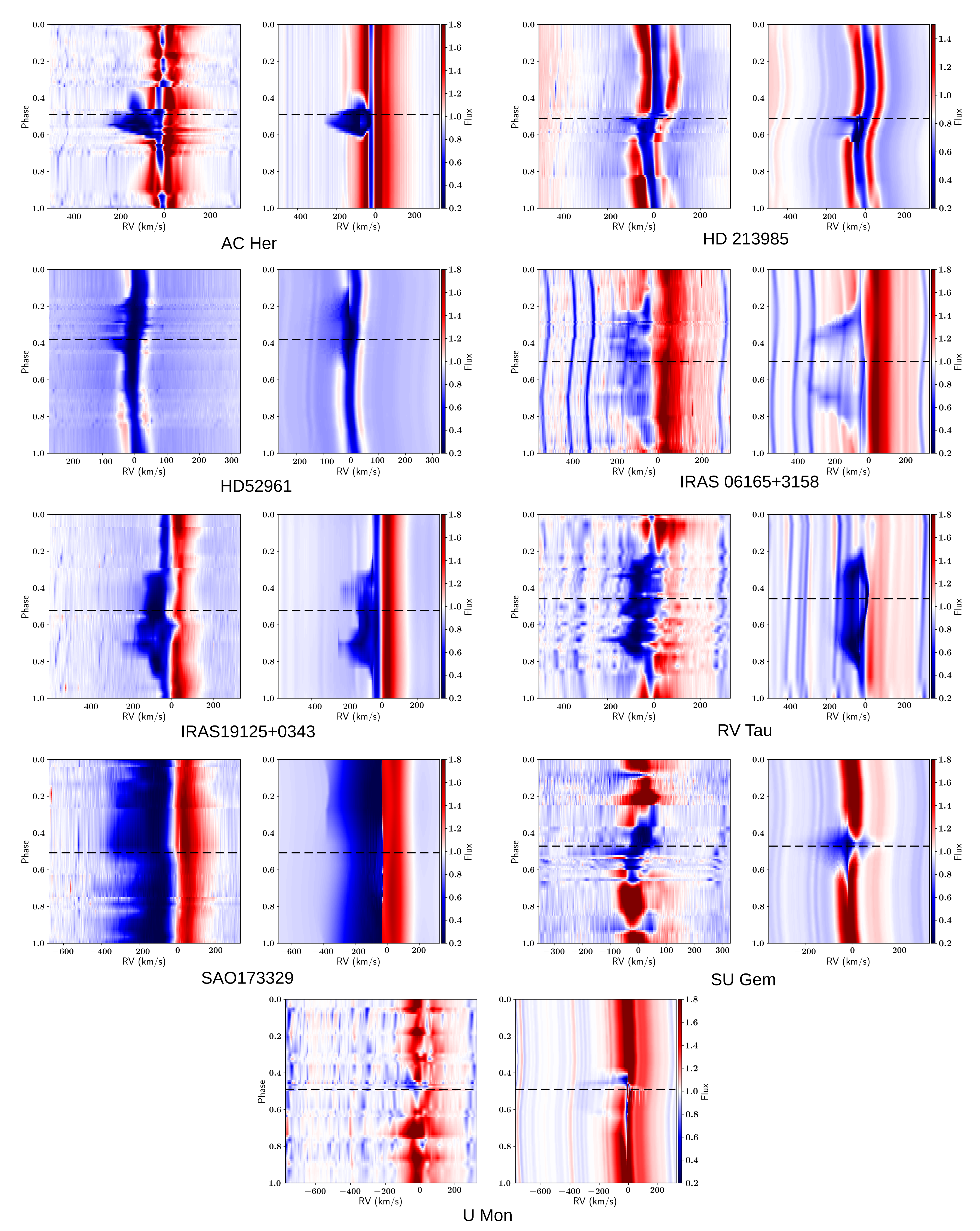}
    	\caption{Dynamic spectra in \halpha\,of AC~Her, HD~213985, HD~52961, IRAS06165+3158, IRAS19125+0343, RV~Tau, SAO~173329, SU~Gem, U~Mon. The observed (\textit{left}) and modelled (\textit{right}) spectra are paired for each object. The spectra are plotted as a function of orbital phase. The velocity on the x-axis denotes the radial velocity shift from the systemic velocity of the binary system. The spectra are normalised to the continuum flux level. The horizontal dashed line marks the phase of superior conjunction.}
    	\label{fig:dynspec1}
    \end{figure*}

 The spectra folded on the orbital period are given in Fig.~\ref{fig:dynspec1}. On the left of each sub-panel, we show the dynamic spectra in the region of the H$_\alpha$ line. We see a large variety in the phase coverage of the jet absorption feature. Another outstanding feature is that the phase of conjunction (indicated by the dashed black line) is not always centred with respect to the jet feature.
             
\section{Methods}\label{sec:methods}
    
\begin{figure*}
    	\centering
    	\includegraphics[width=1\textwidth]{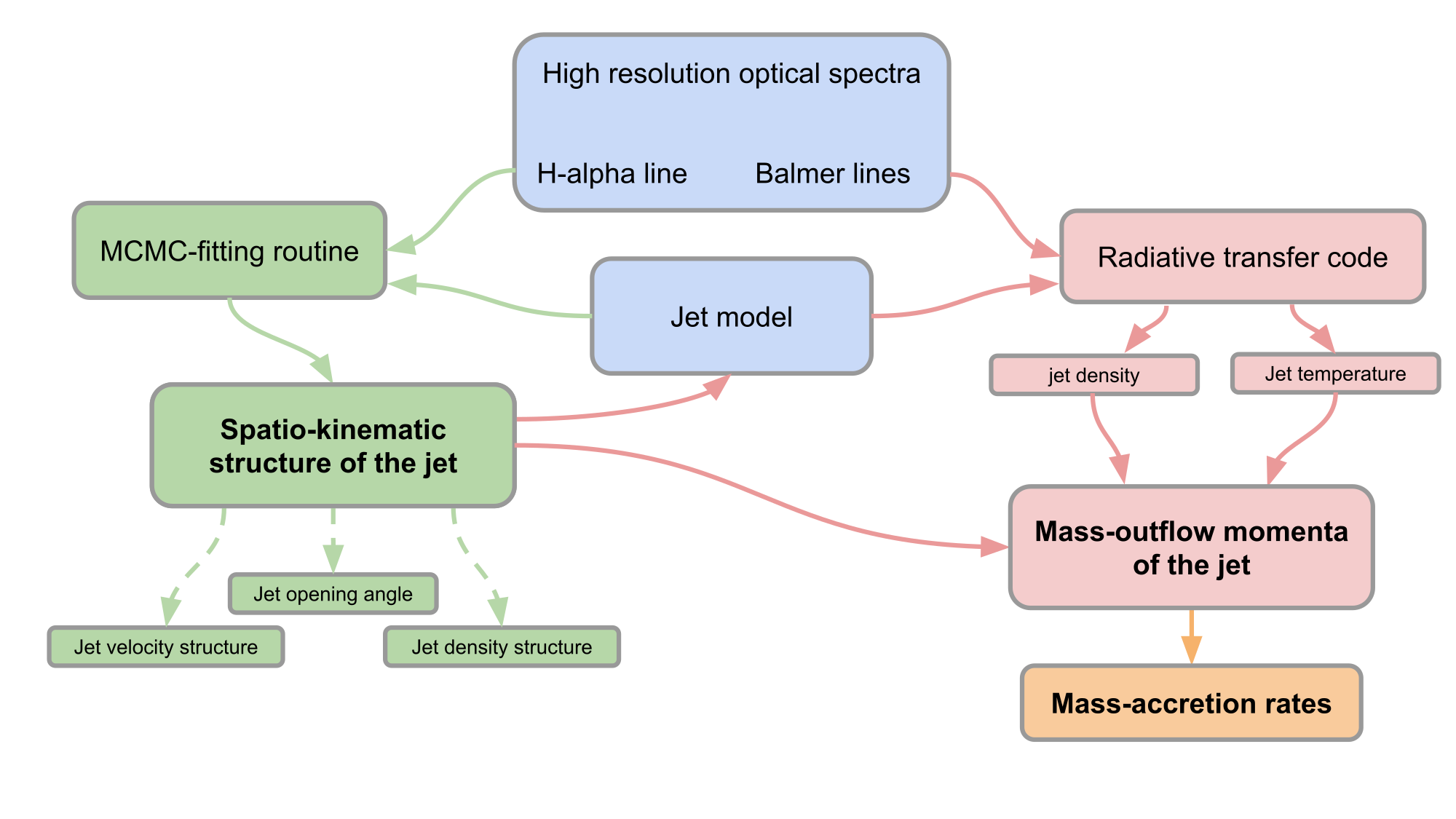}
    	\caption{Flowchart representing the different stages of the  process to determine the spatio-kinematic structure and mass-loss rates of the jet and the mass-accretion rates in the disk. Arrows indicate connections between the different steps. 
    	The first stage consists of the data collection and jet model (\textit{blue}). The second stage (\textit{green}) is the spatio-kinematic modelling, in which the jet model is fitted to the data, using an MCMC-fitting routine. Third is the radiative transfer modelling (\textit{red}). The final stage is the estimation of the mass-accretion rates onto the companion (\textit{orange}).}
    	\label{fig:methodology}
    \end{figure*}

    In this paper, we apply the models and techniques discussed in \cite{bollen19} and Paper~I to the sample of nine objects (see Section~\ref{sec:sample}), in order to determine the spatio-kinematic structure of the jet, its mass loss and the mass-accretion rates onto the companion. 
    A schematic overview of the workflow is shown in Fig.~\ref{fig:methodology}. In short, the process is made up of four main stages: the data collection and jet model description, the spatio-kinematic modelling, the radiative transfer modelling, and finally the determination of the mass-accretion rates. In this section we describe the steps in some detail but refer to the original papers for a more exhaustive description.
    
    The first stage is indicated in blue. We use the high-resolution spectral time series as input data for the modelling. The jet absorption feature is only observed in the Balmer lines. Thus, we only focus on these lines for the modelling. Additionally, we create a jet-binary model, which is used for both the spatio-kinematic modelling and the radiative transfer modelling. This model includes the post-AGB star, the companion star, and its jet. The orbit of the binary is described by the orbital parameters of each object in Tables~\ref{tab:tableorbitalelements} and \ref{tab:tableorbitalelements2}.
    
    In this jet-binary model, we follow the photospheric light that travels from the post-AGB star to the observer. We use a synthetic spectrum from \cite{coelho14} as a model for the photospheric light of the post-AGB star. These spectra are chosen for each object in our sample according to their effective temperature ($T_\text{eff}$), surface gravity ($\log \text{g}$), and metallicity ($[\text{Fe/H}]$) as derived from detailed spectroscopic analyses in the literature. We divide the photospheric light from the post-AGB star up into $N_r=50$ rays. The background spectrum of every orbital phase, is determined by interpolating the dynamic spectra of the orbital phases which are not affected by the jet (see Paper~I and II for a description of this process).
    When these rays travel through the jet, we calculate the radiative transfer through the jet along these rays with $N_j= 2001$ grid points between the point of entry and exit in the jet (see also Fig.~2 of Paper~I). 
    
    Our algorithm includes three possible parametric jet configurations with their own parameters. These configurations are based on three jet- or wind-launching mechanisms: the stellar jet model by \cite{matt05}, the disk-wind model by \cite{blandford82}, and the X-wind model by \cite{shu94}. In the stellar jet model, the matter in the disk is accreted along the magnetosphere of the companion star. Part of this matter will be ejected along the open magnetic field lines of the star, creating the stellar jet. The famous disk wind formalism by \cite{blandford82} describes the magneto-centrifugal ejection of the disk matter above and below the midplane. In this model, a large scale magnetic field is anchored to the disk and rotates at Keplerian velocities. The disk gas flows away along these magnetic field lines, creating the disk wind. The last model, the X-wind, by \cite{shu94} is similar in nature to the disk wind: the disk matter is also ejected magneto-centrifugally. However, in this X-wind, the matter is ejected from a narrow region at the inner disk radius, which is in co-rotation with the star. This is in contrast with the disk wind, where the ejection zone starts at the inner disk region and extends further out. The jet parameters include the jet opening angle, the jet velocity structure, the scaled jet density structure, the inclination angle of the binary system, and the radius of the post-AGB star (see also Table 1 in Paper~I). 
    	
    The second stage of the process is the spatio-kinematic modelling of the jet, indicated in green in Fig.~\ref{fig:methodology}. In this stage, we recreated the absorption features in the \halpha line that occurs due to the absorption of background spectrum by the jet. This is done by first choosing a set of jet parameters in the jet model and then calculating the absorption for that particular jet model. The radiative transfer through the jet is computed at each phase in the orbit for which we have observations. At this stage, we do not calculate the absolute densities in the jet, but rather a scaled density structure $\rho_\text{sc}(\theta, z)$, which is dimensionless and a function of the polar angle in the jet ($\theta$) and the height above the disk plane in the jet ($z$):
    \begin{equation}
        \rho_\text{sc}(\theta, z) = \left( \frac{\theta}{\theta_\text{out}}\right)^p 
                                    \left(  \frac{z}{1\,\text{AU}}\right)^{-2},\label{eq:scaled_density}
    \end{equation}
    with $\theta_\text{out}$ the outer jet angle and the exponent, $p$, a free parameter in the model for which we apply a range from $-$16 to +16. The optical depth through one line element on the ray that travels through the jet is calculated as follows:
    \begin{equation}
        \Delta \tau_\nu(s) = c_\tau\, \rho_\text{sc}(s)\,\Delta s, \label{eq:tau_skm}
    \end{equation}
    
    with $s(\theta, z)$ the location of a line element along the ray that travels through the jet, $\Delta s$ the length of the line element, and $c_\tau$  an opacity-like scaling parameter, which is a free parameter in the model. By including this scaling parameter, the calculation of the optical depths through the jet is fast, which decreases computational time significantly. This allows us to use a Markov-Chain Monte Carlo (MCMC) fitting routine in order to fit these synthetic spectral time series to the observations. In this work, we implement the \texttt{emcee}-package by \cite{foreman13}. The MCMC fitting routine finds the best-fitting model parameters.  A detailed description of the fitting routine is provided in \cite{bollen19} and is publicly available on Github\footnote{https://github.com/bollend/spatiokm}. At this stage of our algorithm, we derived the detailed comparison between the observations and model spectra as depicted in Fig~\ref{fig:dynspec1}.

    The third stage of our analysis procedure is the radiative transfer modelling, indicated in red in Fig.~\ref{fig:methodology}, by which we obtain the absolute density structure of the jet. We start by using the best-fitting jet parameters from the spatio-kinematic modelling to create a jet model. This jet model is then used as input for the detailed radiative transfer modelling. The observables are now the impact of the jet on the Balmer series ($H_\alpha$ to $H_\delta$). We model these first four lines of the Balmer series for a grid of jet temperatures and densities.  Here, we solve the formal solution of the 1D radiative transfer equation numerically. We assume thermodynamic equilibrium and an isothermal jet to compute the excitation of H in the different excitation stages. Thus, the source function $S_\nu$ is described by the Planck function $B_\nu$ \citep[][see Chapter 1]{rybicki} and the intensity along the whole ray through the jet is computed as follows:
    \begin{align}
        \begin{split}
            I_\nu(\tau_{N_j}) \,= \,&I_\nu(\tau_{0})\,\mathrm{e}^{-(\tau_{0}-\tau_{N_j})} + B_\nu(\tau_{N_j})\,\Big[1-\mathrm{e}^{-(\tau_{N_j-1} - \tau_{N_j})}\Big] \\
            &+ \sum\limits_{i=1}^{N_j-1} B_\nu(\tau_{i})\,\Big[1-\mathrm{e}^{-(\tau_{i-1} - \tau_{i})}\Big]\,\mathrm{e}^{-(\tau_{i}-\tau_{N_j})}, \label{eq:rtnum}
        \end{split}
    \end{align}
 with $\tau_i$ being the optical depth at line element $i$. The optical depth is calculated as 
    \begin{equation}
        \tau_{i+1} - \tau_i = \rho_i \,\kappa_{\nu,i}\, \Delta s = \alpha_{\nu,i} \,\Delta s,
    \end{equation}
     with $\kappa_{\nu,i}$ being the opacity and $\alpha_{\nu,i}$ the absorption coefficient. Using the Boltzmann equation, the absorption coefficient can be written as
    \begin{equation}
    \alpha_\nu(s) = \frac{\pi e^2}{m_e c}\,\phi_\nu\, n_l(s)\, f_{lu}\,\Big[1 - \mathrm{e}^{-\Delta E/k_BT_{jet}}\Big],\label{eq:abscoeff} 
    \end{equation}
    with $m_e$ being the electron mass, $\phi_\nu$ the Voigt profile, $n_l$ the number density in the lower energy level, $f_\text{lu}$ the oscillator strength, $\Delta E$ the difference between the upper and lower energy levels, $k_B$ the Boltzmann constant, and $T_{jet}$ the jet temperature.  Next, we compute the equivalent width (EW) of the synthetic lines. These equivalent widths are then fitted to the equivalent width of the observed Balmer lines, in order to find the best-fitting jet density and temperature. By combining the results from the spatio-kinematic modelling (jet structure) and the radiative transfer modelling (absolute jet densities), we can determine the mass-outflow momenta in the jet. The jet is always seen in absorption against the background spectrum which puts strong constraints on the possible temperature. The radiative transfer modelling is described in more detail in Paper~I. The radiative transfer code is publicly available on Github\footnote{https://github.com/bollend/jet\_accretion}.
    
    The fourth and final stage is the estimation of mass-accretion rates onto the companion from which we investigate the source that could feed the accretion disk around the companion: the post-AGB star or the circumbinary disk. This stage is indicated in orange in Fig.~\ref{fig:methodology}. We infer mass-accretion rates onto the companion star from the jet mass-loss rates. We do this by assuming an ejection efficiency ($\dot{M}_\text{jet, tot}/\dot{M}_\text{accr}=0.4$) but a wide range of values are mentioned in the literature \citep[e.g.][]{ferreira06, blackman14}. Part of the accretion disk is also accreted onto the companion star but observationally this fraction is not constrained.
    
    
    In order to determine the main source feeding the accretion disk around the companion, we compare the deduced mass-loss rates by the jet with the estimated mass-transfer rates from the post-AGB star and the circumbinary disk to the companion. The mass accretion and mass loss rate estimation are described in more detail in Section \ref{sec:discussion}. In this part of the analysis, we include the results for the other seven jet-creating post-AGB binaries from Paper~I and II.


\section{Results}\label{sec:results}


    \subsection{The spatio-kinematic modelling}\label{ssec:skm}
    
\begin{figure*}[!h]
 	\centering
 	\begin{subfigure}[b]{.5\linewidth}
 		\centering\large 
 		\includegraphics[width=1\linewidth]{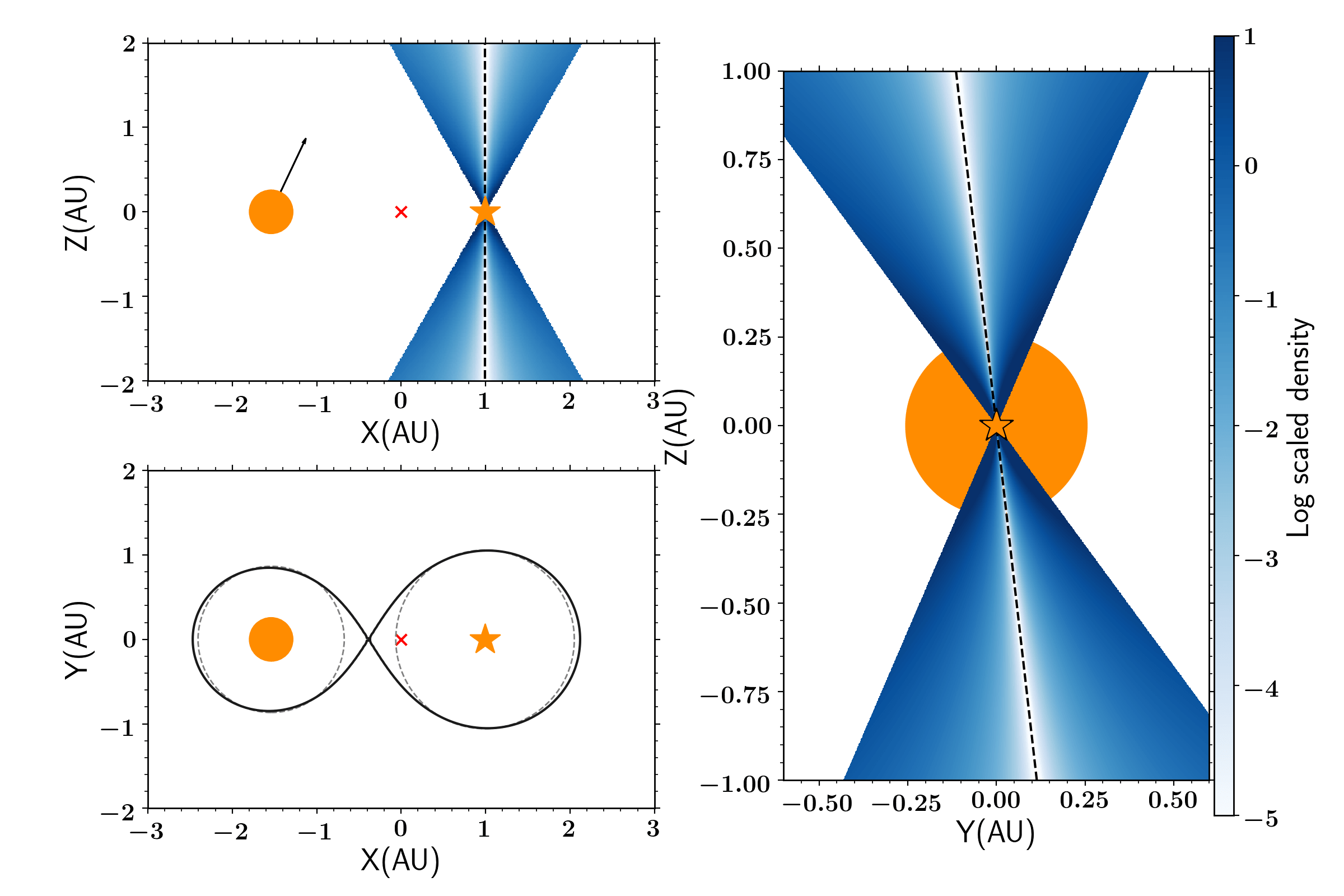}
 		\caption*{(\#1) AC\,Her}\label{fig:dyngeomacac}
 	\end{subfigure}%
 	\begin{subfigure}[b]{.5\linewidth}
 		\centering\large 
 		\includegraphics[width=1\linewidth]{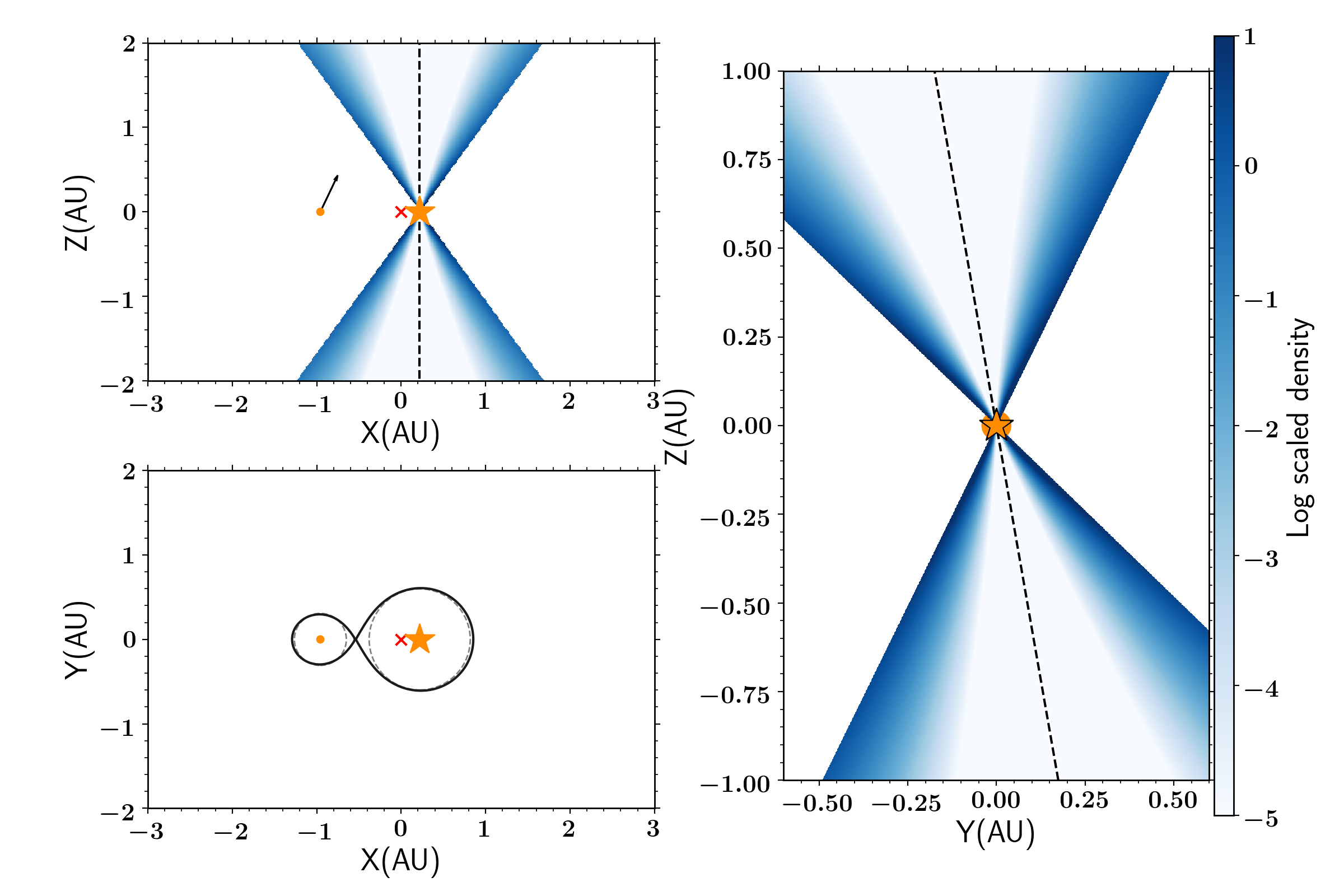}
 		\caption*{(\#2) HD\,213985}\label{fig:dyngeomhd21}
 	\end{subfigure}
 	
 	\begin{subfigure}[b]{.5\linewidth}
 		\centering\large
 		\includegraphics[width=1\linewidth]{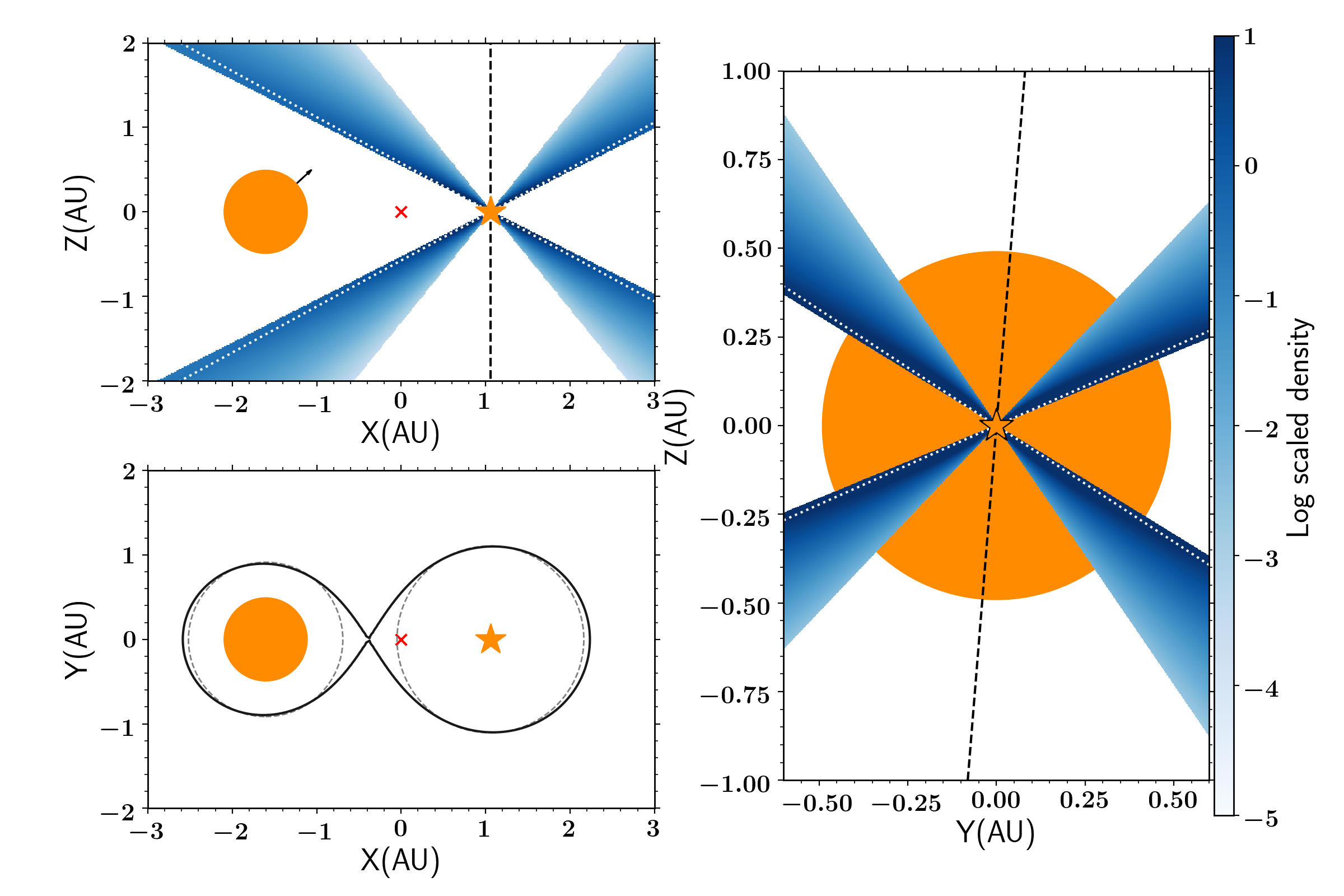}
 		\caption*{(\#3) HD\,52961.}\label{fig:dyngeomhd52}
 	\end{subfigure}%
 	\begin{subfigure}[b]{.5\linewidth}
 		\centering\large
 		\includegraphics[width=1\linewidth]{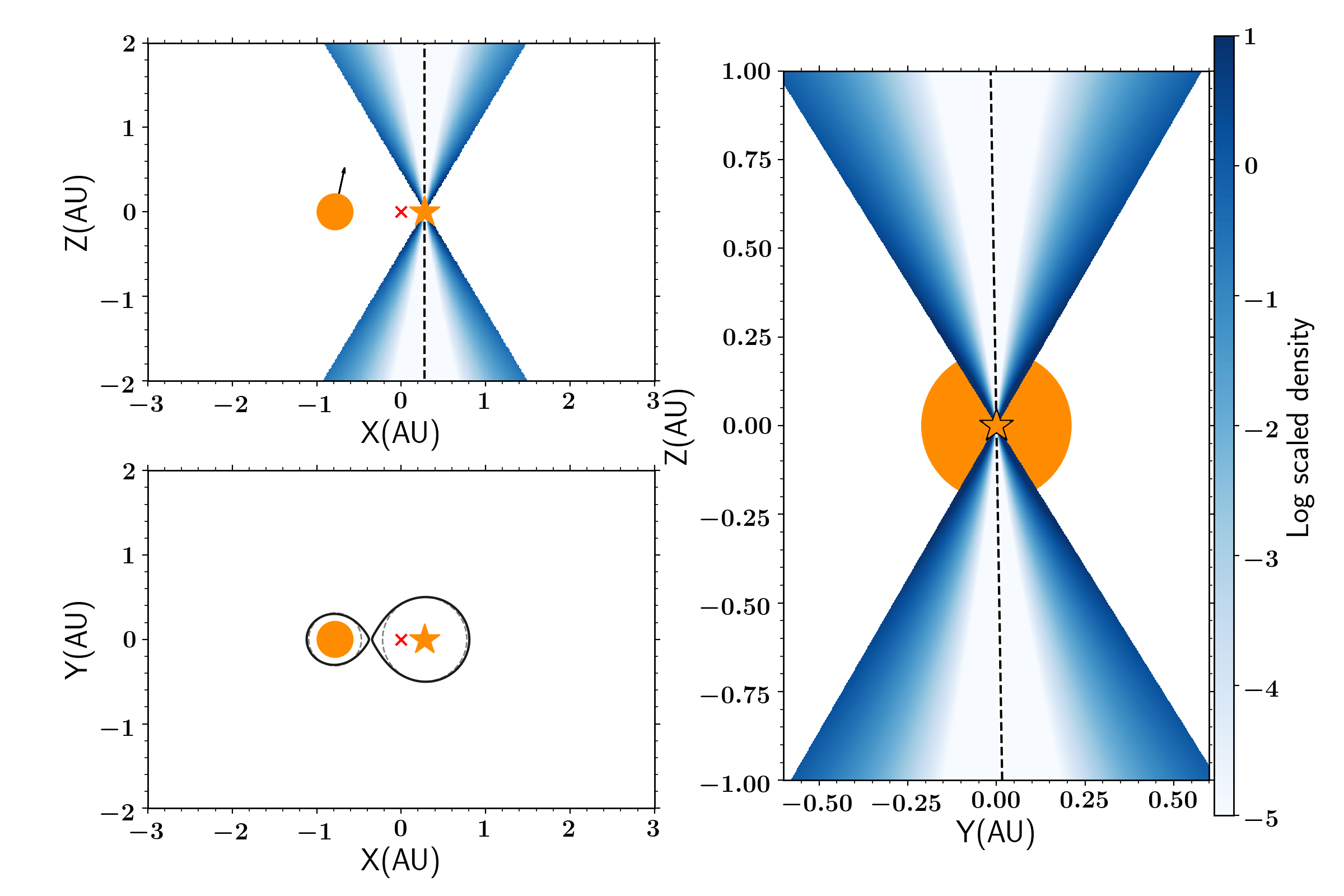}
 		\caption*{(\#4) IRAS06165+3158.}\label{fig:dyngeomiras06}
 	\end{subfigure}
 	
 	\begin{subfigure}[b]{.5\linewidth}
 		\centering\large
 		\includegraphics[width=1\linewidth]{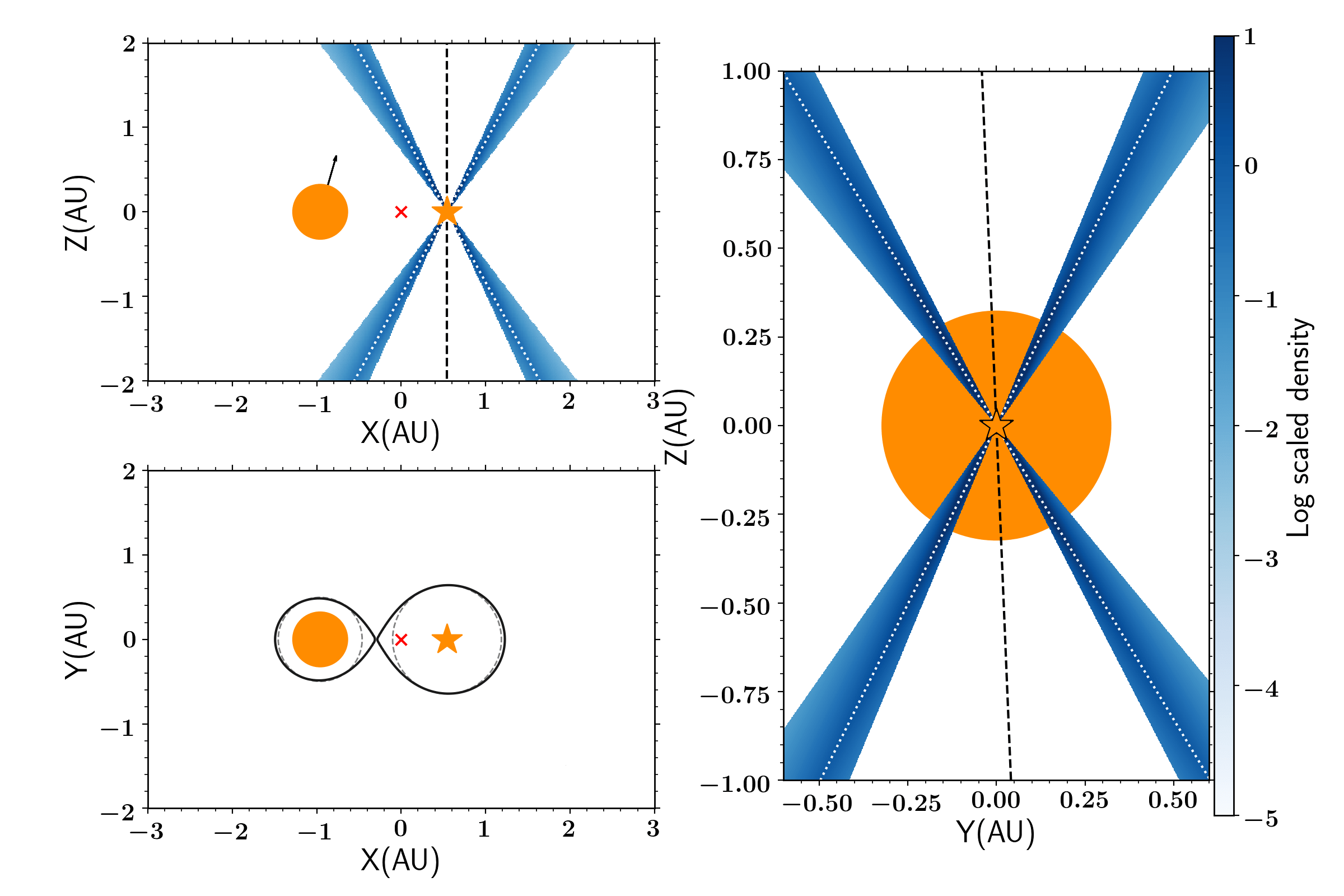}
 		\caption*{(\#5) IRAS\,19125+0343.}\label{fig:dyngeomiras19125}
 	\end{subfigure}%
 	\begin{subfigure}[b]{.5\linewidth}
 		\centering\large
 		\includegraphics[width=1\linewidth]{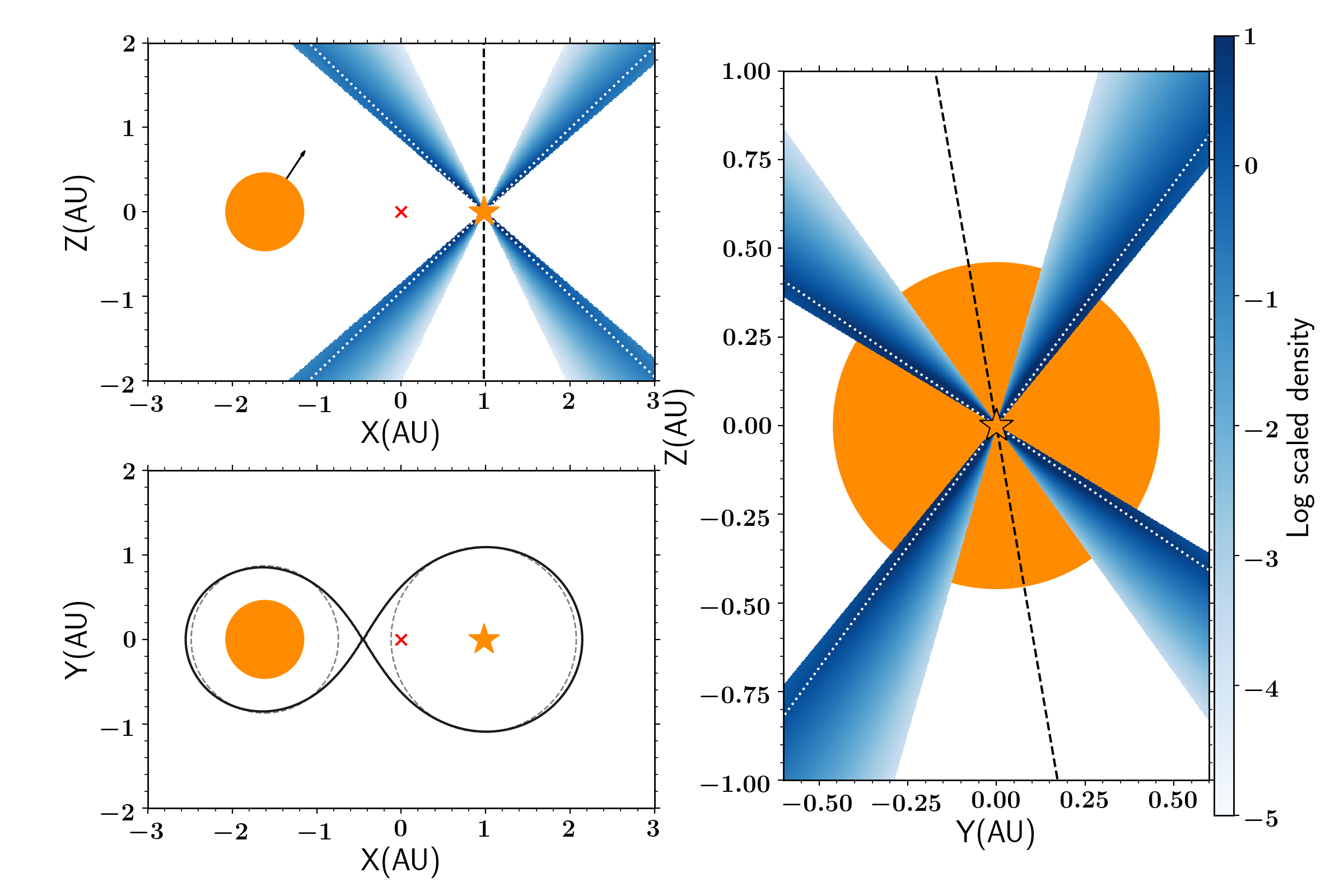}
 		\caption*{(\#6) RV\,Tau.}\label{fig:dyngeomrv}
 	\end{subfigure}
 	\caption{Geometry of the binary system and the jet in six objects in this sample. The plots show the system during superior conjunction, when the post-AGB star is directly behind the companion, as viewed by the observer. The plot in the upper-left panel shows the system viewed edge-on. The arrow gives the line of sight towards the observer. The right-hand-side panel shows the system viewed edge-on and during superior conjuction. The lower-left panel shows the binary system viewed pole-on. The different elements in the plots are the post-AGB star (\textit{orange circle}), the location of the companion (\textit{orange star}), the centre of mass (\textit{red cross}), the jet axis (\textit{dashed black line}) and inner jet edges (\textit{dotted white lines}), and the Roche lobes (\textit{full black line}).}\label{fig:dyngeom}
 \end{figure*}

\begin{figure*}[!h]
 	\begin{subfigure}[b]{.5\linewidth}
 		\centering\large
 		\includegraphics[width=1\linewidth]{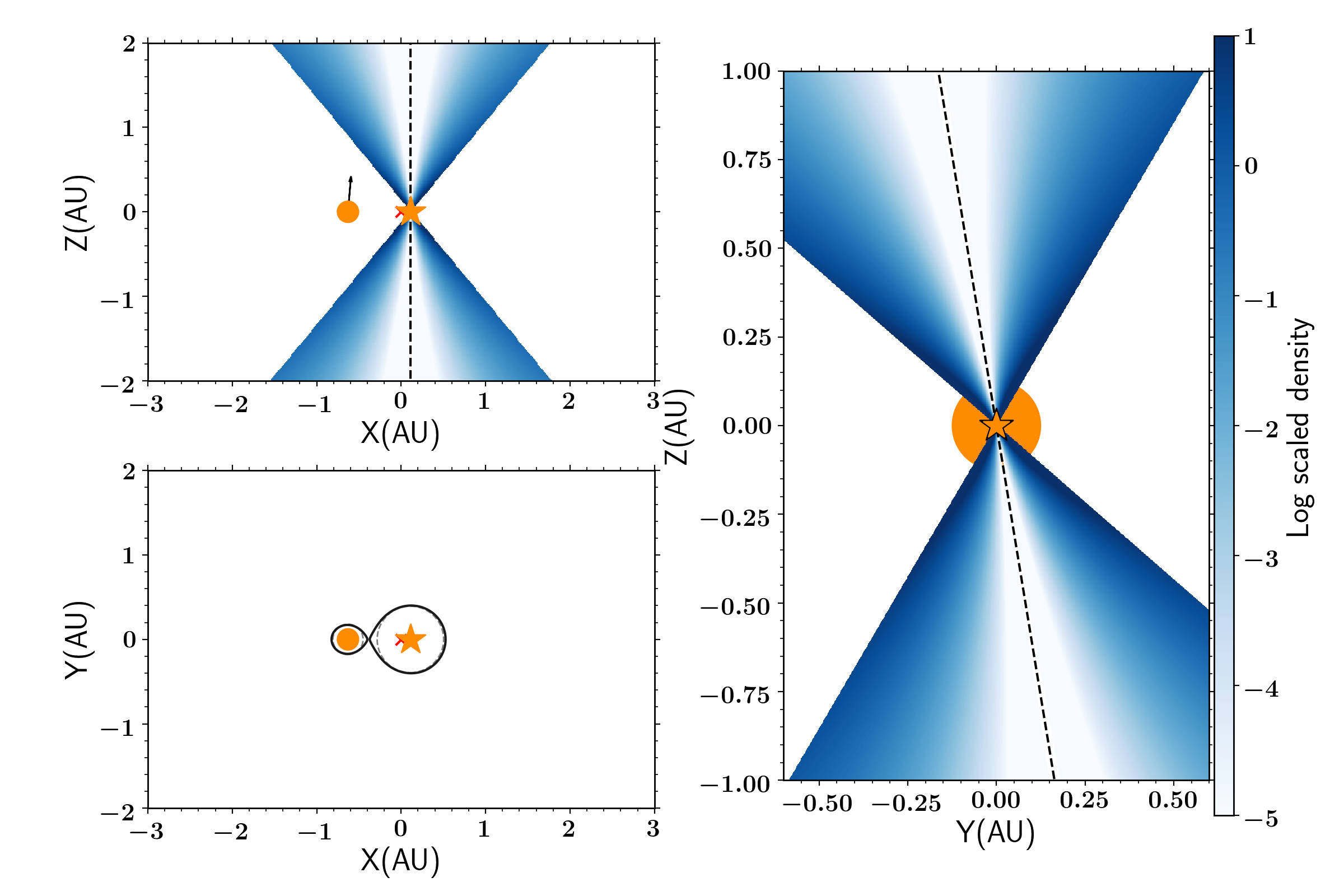}
 		\caption*{(\#7) SAO\,173329.}\label{fig:dyngeomsao}
 	\end{subfigure}%
 	\begin{subfigure}[b]{.5\linewidth}
 		\centering\large
 		\includegraphics[width=1\linewidth]{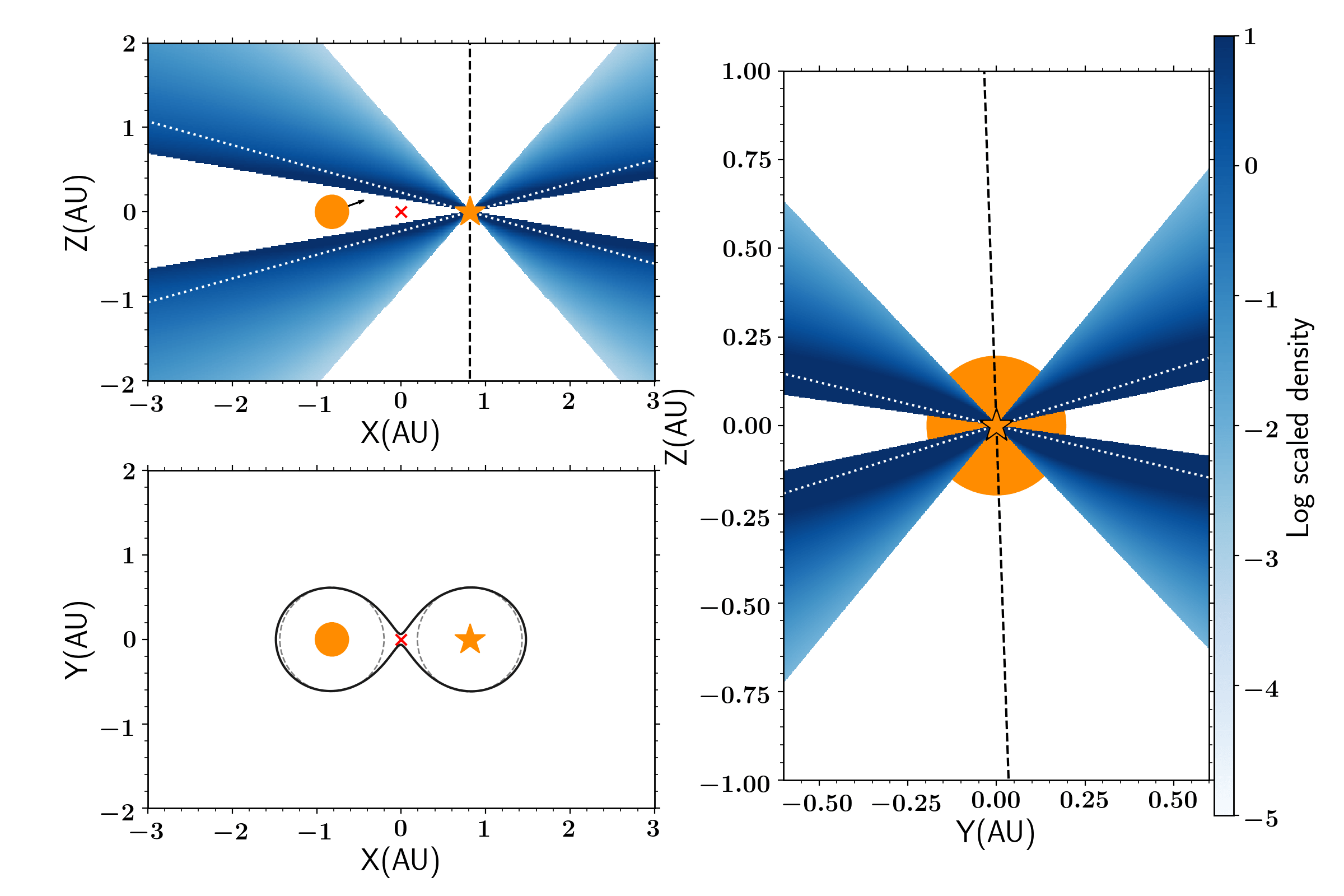}
 		\caption*{(\#8) SU\,Gem.}\label{fig:dyngeomsu}
 	\end{subfigure}
 	
 	\begin{subfigure}[b]{.5\linewidth}
 		\centering\large
 		\includegraphics[width=1\linewidth]{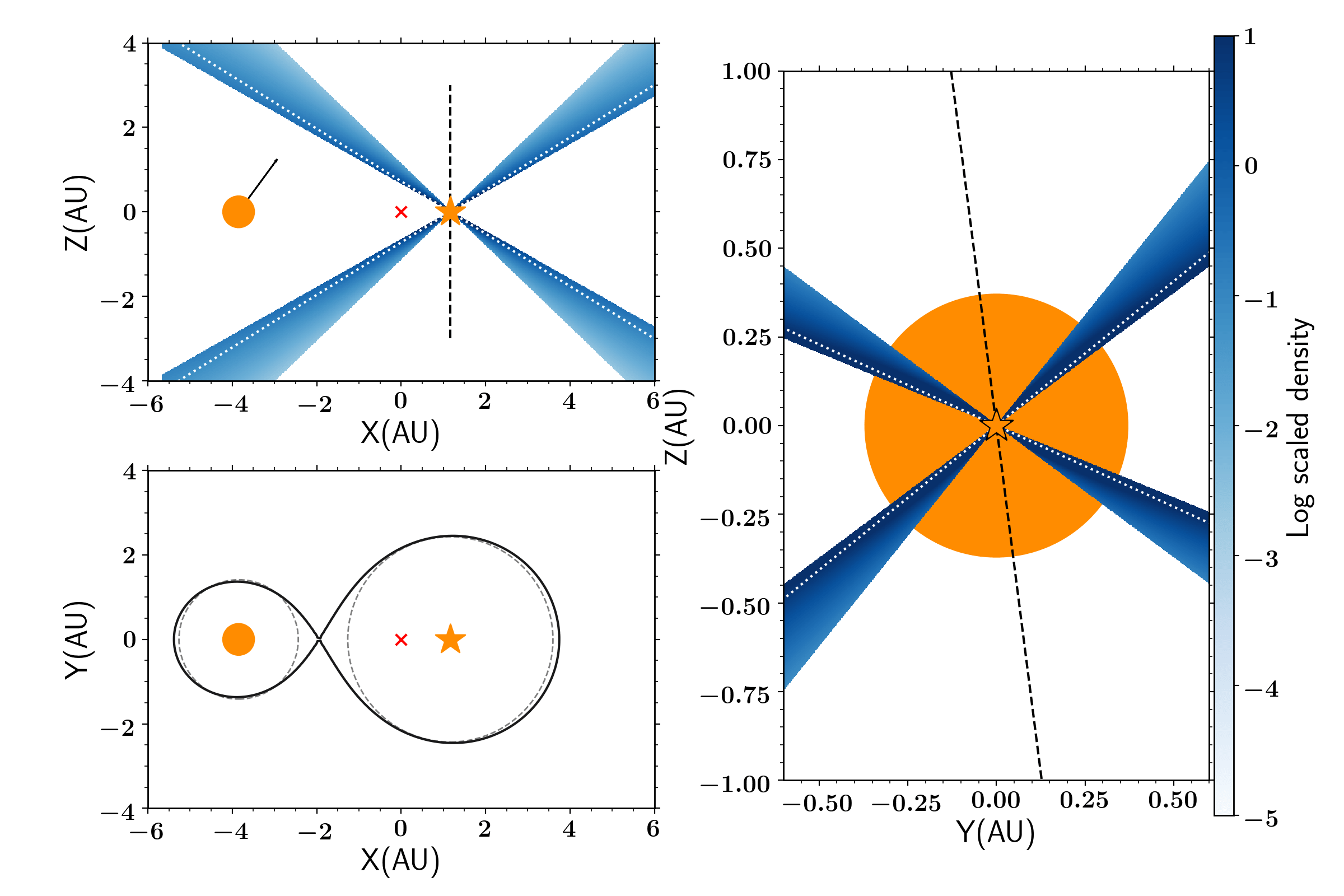}
 		\caption*{(\#9) U\,Mon.}\label{fig:dyngeomumon}
 	\end{subfigure}
 	\caption{Similar as Figure~\ref{fig:dyngeom}, but for SAO\,173329, SU\,Gem, and U\,Mon.}\label{fig:dyngeom_updated}
 \end{figure*}

         The comparison between the observed and modelled dynamic spectra of the spatio-kinematic fitting are shown in Fig.~\ref{fig:dynspec1}. The best-fitting jet configurations and parameters are presented in Table~\ref{tab:ap_skmresults1}. The geometries are presented in Figs.~\ref{fig:dyngeom} and~\ref{fig:dyngeom_updated} which give a good representation of the geometry and density structure of the systems.

     
        Fig.~\ref{fig:dynspec1} and the reduced chi-square values (see Table~\ref{tab:ap_skmresults1}) show that the modelling has been successful in reproducing the observations. The variety of spectral behaviour in  \halpha\, is large, but the modelling is versatile enough to accommodate all the different objects. The duration of the absorption feature varies between $10\%$ and $100\%$ of the orbital period. Hence, for some objects, such as HD\,213985, the absorption feature is observed during a very short period of time, whilst for other objects, such as SAO~173329, the absorption feature is always observable. Other observational characteristics of this absorption feature, including its blue-shifted extent, its absorption strength, and the variation in this feature across the orbit, are all well reproduced by the modelling. Figs.~\ref{fig:dyngeom} and~\ref{fig:dyngeom_updated} illustrate also the line-of-sight towards the system which can be related to the phase coverage as given in Fig.~\ref{fig:dynspec1}.
        
        Our sample includes systems that are classified as RV Tauri pulsators \citep{wallerstein02, manick17, manick19}. The model performs obviously better for the non-pulsating objects ($\chi^2_\nu < 2.5$) than for pulsating objects. AC\,Her, RV\,Tau, SU\,Gem and U\,Mon, have reduced chi-square values ranging from $3.5$ to $9$. This is because the model does not account for the added complexity of strong shock-emission in the \halpha\,lines of these pulsating stars. Despite the shock features in the spectra, the spatio kinematic model can still reproduce the observed jet absorption feature in \halpha\,for these strong pulsators.
        
    The jet opening angle, the binary inclination, the jet orientation, the jet velocity structure, the jet density structure, and the post-AGB stellar radius are free parameters in our jet model. Also the post-AGB radius is assumed to be a free parameter because the distances to the objects, and hence their luminosities, are poorly constrained. Below, we compare the results of the spatio-kinematic modelling of all nine objects and discuss how the jet and binary properties impact the observed quantities.

        \begin{sidewaystable*}[]
        	\begin{center}
        		\caption{Best-fitting jet configuration and jet parameters for the spatio-kinematic model for the objects in our sample. }
        		\label{tab:ap_skmresults1}
        		\begin{tabular}{l c c c c c c c c c}
                \hline \hline
                Object                             &  AC Her           & HD\,213985        &   HD\,52961       & IRAS06165+3185   &  IRAS19125+0343  & RV\,Tau          & SAO~173329      &  SU\,Gem         &  U Mon \\
                Parameter                          &                   &                   &                   &                  &                  &                  &                 &                  &        \\
                \hline
                Best-fit configuration             &  stellar jet      &  stellar jet      &    X-wind         &  stellar jet     &  X-wind          &  X-wind         &  stellar jet     &  X-wind          &  X-wind\\
                $i$ ($\degr$)                      &  $49.8\pm0.3$     &  $50.1\pm0.3$     &   $70.6\pm0.6$    &  $28.6\pm0.1$    &  $39.1\pm0.2$    &  $62.4\pm0.5$   &  $12.1\pm0.1$    &  $81.9\pm0.2$    &  $61.4\pm0.2$ \\
                $\theta_\text{out}$ ($\degr$)      &  $29.8\pm0.1$     &  $36.2\pm0.3$     &   $68.7\pm0.9$    &  $31.0\pm0.1$    &  $37.4\pm0.3$    &  $50.3\pm0.6$   &  $39.6\pm0.3$    &  $79.8\pm0.3$    &  $60.4\pm0.2$\\
                $\theta_\text{in}$ ($\degr$)       &                   &                   &   $60\pm2$        &                  &  $34.2\pm0.4$    &  $36.5\pm0.6$   &                  &  $74.3\pm0.7$    &  $58.1\pm0.6$\\
                $\theta_\text{cav}$ ($\degr$)      &                   &                   &   $28.9\pm9$      &                  &  $25.0\pm0.1$    &  $29.2\pm0.4$   &                  &  $41.6\pm0.9$    &  $46.1\pm0.4$ \\
                $\phi_\text{tilt}$ ($\degr$)       &  $6.5\pm0.1$      &  $9.9\pm0.2$      &   $-1.3\pm0.6$    &  $0.93\pm0.03$   &  $2.2\pm0.1$     &  $12.2\pm0.1$   &  $9.3\pm0.2$     &  $1.97\pm0.09$   &  $7.3\pm0.2$ \\
                $v_\text{in}$ (km\,s$^{-1}$)       &  $369\pm3$        &  $600\pm20$       &   $206\pm9$       &  $402\pm2$       &  $176\pm2$       &  $352\pm5$      &  $489\pm4$       &  $405\pm13$      &  $421\pm8$ \\
                $v_\text{out}$ (km\,s$^{-1}$)      &  $80\pm2$         &  $33\pm1$         &   $1.1\pm0.3$     &  $38.4\pm0.2$    &  $59\pm1$        &  $54\pm2$       &  $37.2\pm0.2$    &  $0.9\pm0.4$     &  $5.1\pm0.6$ \\
                $p_\text{v}$                       &  $0.76\pm0.10$    &  $1.2\pm0.1$      &   $1.4\pm0.2$     &  $-2.99\pm0.02$  &  $-0.71\pm0.09$  &  $2.53\pm0.05$  &  $-0.67\pm0.05$  &  $1.5\pm0.1$     &  $2.4\pm0.1$ \\
                $p_{\rho\text{,in}}$               &                   &                   &   $15.8\pm0.3$    &                  &  $9.2\pm0.3$     &  $0.4\pm0.2$    &                  &  $10.1\pm0.6$    &  $14.9\pm0.3$\\
                $p_{\rho\text{,out}}$              &  $3.1\pm0.1$      &  $14.7\pm0.6$     &   $-0.0\pm0.5$    &  $9.6\pm0.1$     &  $14.9\pm0.2$    &  $9.3\pm0.3$    &  $6.9\pm0.2$     &  $14.8\pm0.9$    &  $13.9\pm0.4$\\
                $c_\tau$                           &  $2.88\pm0.02$    &  $2.45\pm0.05$    &   $1.39\pm0.06$   &  $2.49\pm0.02$   &  $2.27\pm0.03$   &  $1.22\pm0.05$  &  $3.02\pm0.02$   &  $0.32\pm0.08$   &  $2.06\pm0.09$\\
                $R_\text{1}$ (R$_\odot$)           &  $54.8\pm0.9$     &  $8.6\pm0.3$      &   $131\pm4$       &  $45.2\pm0.3$    &  $60\pm2$        &  $56.3\pm0.9$   &  $26.7\pm0.3$    &  $41.9\pm0.7$    &  $79.6\pm0.7$\\
                $\chi^2_{\nu}$                     &  $3.6$            &    $1.9$          &    $1.3$          &   $2.6$          &  $1.1$           &   $9.9$         &  $2.5$           &  $3.6$           &  $3.9$\\
                \hline
        		\end{tabular}
        		\tablefoot{The tabulated parameters are: inclination angle of the binary system $i$, jet outer angle $\theta_\text{out}$, jet inner angle $\theta_\text{in}$, jet cavity angle  $\theta_\text{cav}$, jet tilt $\phi_\text{tilt}$ , inner jet velocity  $v_\text{in}$, jet velocity at the jet edges  $v_\text{out}$, exponent for the velocity profile  $p_\text{v}$, exponent for the density profile for the outer and inner region $p_{\rho \text{,out}}$ and  $p_{\rho\text{,in}}$, optical depth scaling parameter  $c_\tau$, the radius of the post-AGB star $R_\text{1}$, and the reduced chi-square for the jet configuration. The objects with a stellar jet configuration only have an outer jet angle $\theta_\text{out}$ and one exponent for the density profile  $\theta_\text{out}$. The errors are the 1$\sigma$ uncertainty interval and are statistical only. }
        	\end{center}
        	
        \end{sidewaystable*}

        \begin{itemize}
        
        	\item \textbf{Jet half-opening angle and binary inclination:\\} 
        	The jet half-opening angle and binary inclination angle have a direct effect on the duration of the absorption feature. As the jet half-opening angle increases, our line of sight goes through the jet for a longer period of time. The orbital inclination angle has the opposite effect. As the inclination angle of the binary system increases, and the system is observed more edge-on, our line of sight towards the background AGB star traverses the jet closer to its base, where the jet is narrower and the jet will be in the line of sight for a shorter period of time. We refer to the duration of the absorption feature as the phase coverage of the absorption, so in terms of the orbital period.
        	
        	 Different combinations of the jet opening angle and the orbital inclination can result in similar absorption line phase coverage.  We investigate this effect here. SAO~173329 is a good example of a system in which jet absorption is always present. The inclination angle in this system is significantly smaller than the jet half-opening angle. Thus, the line of sight from the post-AGB star to the observer always passes through the jet, irrespective of the orbital phase. Other systems with a large phase coverage of the absorption are IRAS06165+3158 and IRAS19125+0343. These objects also have an inclination angle that is smaller than the jet half-opening angle. For AC~Her, HD~213985  HD~52961, SU~Gem, and U~Mon, the phase coverage is $50\%$ or smaller, which is explained by a positive difference between inclination and jet half-opening angle. The only outlier in this sample is RV\,Tau. The best-fitting model for this object has an inclination angle that is larger than the jet half-opening angle, yet the absorption is observed during $70\%$ of the orbital phase. The jet tilt and the large post-AGB radius makes that the absorption is indeed observable for a longer period of time. We use the MCMC posterior density distributions for the determination of the errors on the model parameters as well as the detection of eventual local minima. 
        	
        	\item \textbf{Jet tilt:\\} A jet that is tilted with respect to the orbital axis of the binary system can cause a shift in orbital phase of the absorption, such that it is no longer centred on the time of superior conjunction. If the detected cone of the jet is tilted away from the direction of travel of the companion, the occultation of the post-AGB star by the jet will occur later in the orbital phase. This effect is clearly present in the dynamic spectra of AC~Her. The absorption feature occurs later in the orbital phase. Consequently, these systems have a positive jet tilt between $5\degr$ and $15\degr$, which corresponds to an occulting jet lobe that is tilted away from the direction of travel of the companion. For HD~52961, the absorption feature peaks before superior conjunction. This object has a negative jet tilt, corresponding to a jet that is tilted towards the direction of travel. Three objects (IRAS06165+3158, IRAS19125+0343, and SU~Gem) in our sample have a small to no jet tilt ($<4\degr$).
        	
        	The jet tilts could be explained by a precessing motion of the jet, which is often observed in jets launched from binary systems \citep{bollen21, choi17, sahai17}. Jet precession would be mainly caused by a tilted accretion disk around the companion.  Alternatively, the orbital motion of the binary and the Coriolis effect could also have a significant dynamical influence on the outflow \citep{fendt98, tuthill08,boschramon16}. Due to the orbital motion, the gas seen in the jet will have been launched at different orbital phases. This could eventually result in the bending of the jet and would result, on larger scales, in a helical pattern. The degree by which the helical pattern coils are wound will depend mainly on the relationship between the jet velocity and the orbital velocity. Once the orbital velocity is $\gtrsim 5\%$ of the jet velocity, the jet deflection can become significant. The outflow velocities at the edges in the jets range from $1\,$\kms to $80\,$\kms, which is on the same order of magnitude as the orbital speed of the companion star. Thus, the jet deflection could become significant in these regions. Moreover, since this jet deflection has a different effect in the slower moving gas compared to the faster moving gas in the jet, these regions could eventually become entangled, which would result in a far more complex geometry of the whole outflow. 
        
        	\item \textbf{Jet velocity structure:\\} The extent of the blue-shifted absorption feature is governed by the fastest outflow velocities of the jet, and ranges from $150\,$\kms to $400\,$\kms in projection. The maximum physical, deprojected jet velocities as constrained by the model are slightly higher: between $150\,$\kms\, and $640\,$\kms.
        	
        	The jet velocities found by the model can be used to constrain the nature of the jet-launching object (the companion in these systems), since the highest jet velocities are on the order of the escape velocity of the jet-launching object \citep{livio99}. The two most plausible candidates for the companion are either a main sequence (MS) star or a white dwarf (WD). Using our model, we are able to rule out compact objects in these systems, since the escape velocity of a WD ($\sim 5000\,$\kms) is about one order of magnitude larger than our deprojected jet velocities. The jet velocities we derive are all around the escape velocity of a MS star ($100-1000\,$\kms), which is a strong indication that the companions in all these systems are MS stars and that we have no objects with a WD companion.
        	
        	\item \textbf{Jet density structure:\\} The jet density structure dictates the strength of the absorption and its variability during orbital motion. A higher density results in a stronger absorption feature. The model predicts that the jet density is higher at the jet's outer surface and lower along the jet's axis. One of our interesting results is a clear model preference for a jet that includes a cavity, defined as a region with very low density. This jet cavity is included naturally in the X-wind and disk wind configurations. In these systems, the central regions of the jet are devoid of material and the bulk of the mass is ejected at angles $\theta > 30\degr$.

        	A jet cavity is not explicitly defined in the stellar jet configuration. The inner density structures for the jets in HD~213985, IRAS06165+3158, and SAO~173329, closely resemble a jet cavity, however, even though the best-fitting configuration is found to be a stellar jet. In these systems, the jet density has a large gradient as a function of polar angle in the jet. Hence, the inner regions in the jet ($\theta < 20\degr$) have extremely low densities (see colour codes in Figs.~\ref{fig:dyngeom} and \ref{fig:dyngeom_updated}), which mimics the jet cavity of the X-wind and disk wind models.
        	
        	\item \textbf{Radius (and luminosity) of the post-AGB star:\\} The final main parameter that directly affects the jet absorption feature is the radius of the post-AGB star. A larger radius prolongs the duration of the jet absorption feature in the time series and increases the jet volume  through which post-AGB light passes on its way to us. The stellar radii found by the model range between $9\,$R$_\odot$ and $100\,$R$_\odot$. 
        	
        	We note that the upper limit of the post-AGB radius was set to $80\%$ of the radius of its Roche lobe. This upper limit was imposed since these post-AGB stars do not show any ellipsoidal variations \citep{kiss07}, which typically appear in stars that fill more than $80\%$ of their Roche lobe \citep{wislon76}.
        	
        	Since the effective temperature of the post-AGB star is derived from the literature (see Table~\ref{tab:tableorbitalelements2}), we can calculate the luminosity of the post-AGB star from it and the stellar radius derived by the model. These luminosities can also be derived from the GAIA-DR2 parallaxes \citep{gaia16,gaia18} in combination with the SED and its integral. The GAIA-DR2 parallax ($\varpi$) is converted to distance using $d=1/\varpi$. The luminosity is then determined using the integral of the observed SED as total flux measurement. We note that the total line of sight reddening is an important correction factor. We estimate the error on the luminosity by propagating the error on the parallax. 
        	
        	We compare the luminosities derived from our model with those calculated from the GAIA parallaxes in Table~\ref{tab:tableorbitalelements2} and Fig.~\ref{fig:luminosities}, which also includes the results from Paper~I and II. In general, the luminosities calculated from the model (y-axis) agree fairly well with those found from the GAIA parallaxes (x-axis). One needs to consider that the GAIA-DR2 release did not take binarity into account. The orbital motion induces a angular displacement which is comparable to the parallax itself which means that for the moment the astrometric distances are still uncertain as also the GAIA DR3 release did not flag and resolve these systems as astrometric binaries. Most objects are too far away for the current quality assessment procedure. More GAIA visits are still required to obtain the astrometric orbits. The luminosities obtained via GAIA data and via our spectroscopic fits differ by less than a factor of four (except for HD\,52961, which differs by a factor of six). Since luminosity is a function of stellar radius squared, this implies that the stellar radius found by the model is correct within a factor of two. We need to wait to the astrometric binary orbits to obtain astrometric distances which will provide a much better constraint for the radii of these objects.
        	
        \end{itemize}
        
        \begin{figure}[h!]
        	\centering
        	\includegraphics[width=1\linewidth]{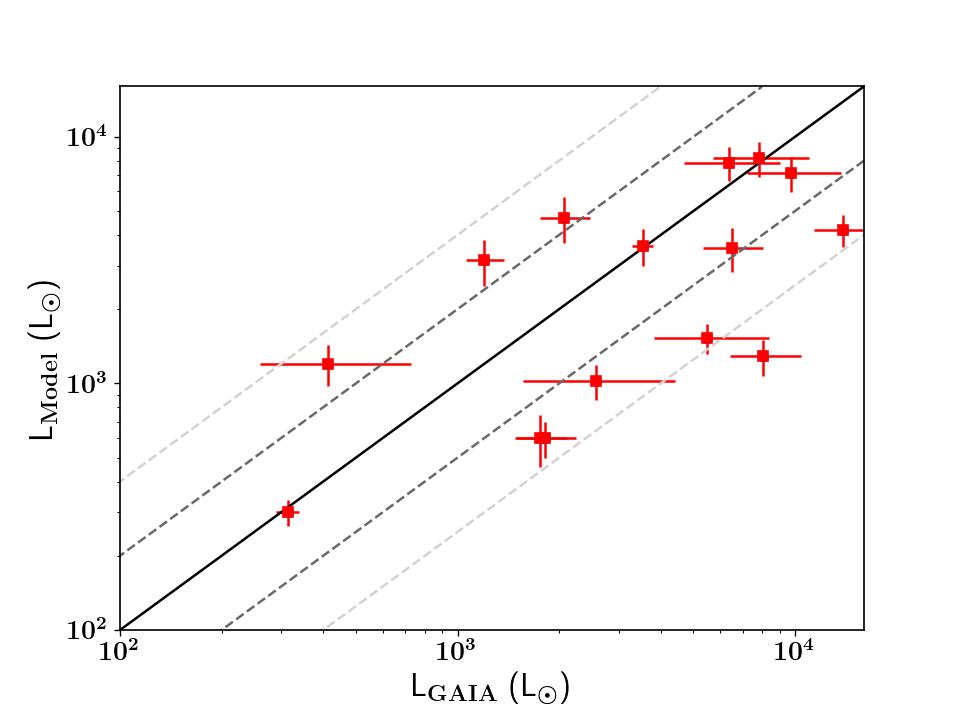}
        	\caption{Comparison of the luminosities derived from the stellar radii found by our spatio-kinematic modelling (y-axis) and those derived from the GAIA parallaxes (x-axis). The black line indicates the unity line (L$_\text{\rm model}=$L$_\text{GAIA}$). The dashed grey lines indicate where L$_\text{\rm model}=2\,$L$_\text{GAIA}$ and L$_\text{\rm model}=0.5\,$L$_\text{GAIA}$. The dashed light grey lines indicate where L$_\text{\rm model}=4\,$L$_\text{GAIA}$ and L$_\text{\rm model}=0.25\,$L$_\text{GAIA}$. }
        	\label{fig:luminosities}
        \end{figure}


    \subsection{Radiative transfer modelling}\label{ssec:rtm}
         \begin{table} 
 	\begin{center}
 		\caption{Best-fitting jet temperature and density from the radiative transfer modelling. The given densities represent the density at the edge of the jet at a height of 1\,AU from the jet base.}
 		\label{tab:rtresults}
 		\begin{tabular}{ll l l}
 			\hline \hline
 			\#  &  Object     & $T_\text{jet}$ & $\log\,n$  \\
 			&   & K              &    m$^{-3}$       \\
 			\hline
 			1  &  AC\,Her    & $5200^{+300}_{-500}$             & $15.8^{+1.4}_{-1}$       \\
 			2  &  HD\,213985 & $5700^{+700}_{-700}$             & $15.8^{+0.8}_{-0.5}$       \\
 			3  &  HD\,52961  & $5100^{+600}_{-600}$             & $15.8^{+1.1}_{-1.1}$       \\
 			4  &  IRAS06135+3158    & $3900^{+200}_{-900}$            & $19.0^{+1.2}_{-1.2}$       \\
 			5  &  IRAS19125+0343    & $5700^{+1200}_{-250}$           & $15.9^{+0.8}_{-0.5}$       \\
 			7  &  SAO~173329  & $5400^{+500}_{-500}$             & $16.7^{+0.6}_{-0.4}$       \\
 			9  &  U\,Mon     & $5000^{+100}_{-600}$             & $18.4^{+0.8}_{-0.2}$       \\
 			\hline
 		\end{tabular}
 	\end{center}
 \end{table}

        In the next step, we use the results from the spatio-kinematic modelling as input for the radiative transfer modelling. We create synthetic Balmer lines for a grid of jet temperatures and densities. The density distribution is a result of the modelling and we define the absolute density at 1 AU on the bipolar symmetry axes and at the edge of the jet cone. Next, we fit the equivalent widths of the synthetic spectra to the observations, in order to find the best-fitting jet temperature and density. The absorption strength in different Balmer lines and their ratios vary according to the chosen temperature and density. For this reason, we include the \hbeta, \hgamma, and \hdelta\,lines in the fitting. For IRAS06165+3158 and U~Mon, we do not include the \hgamma\,and \hdelta\,lines in the fitting due to their low signal-to-noise.
        
        The spectra of RV~Tau and SU~Gem are highly affected by shock-emission features and have low signal-to-noise ratio in \hbeta, \hgamma\ and \hdelta. Hence, we exclude these two objects from our radiative transfer modelling.
        
\begin{figure*}
        	\centering
        	\begin{subfigure}[b]{.33\linewidth}
        		\centering\large 
        		\includegraphics[width=1\linewidth]{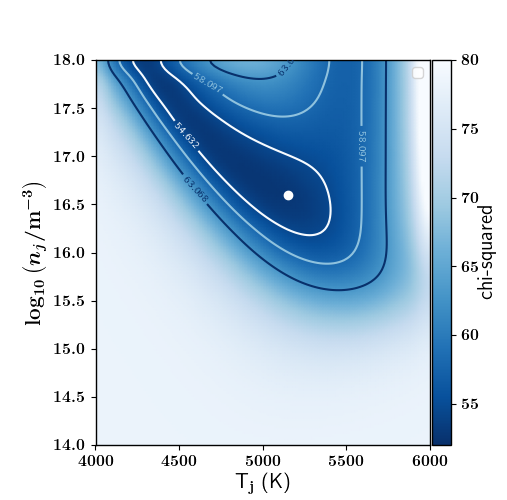}
        		\caption*{(\#1) AC\,Her.}\label{fig:ap_fitrt_chisq_ac}
        	\end{subfigure}%
        	\begin{subfigure}[b]{.33\linewidth}
        		\centering\large 
        		\includegraphics[width=1\linewidth]{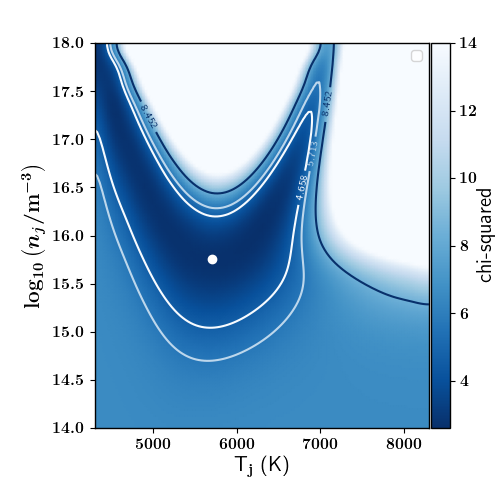}
        		\caption*{(\#2) HD\,213985}\label{fig:ap_fitrt_chisq_hd21}
        	\end{subfigure}%
        	\begin{subfigure}[b]{.33\linewidth}
        		\centering\large 
        		\includegraphics[width=1\linewidth]{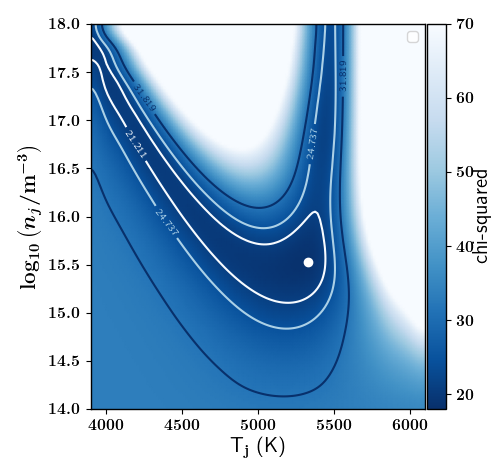}
        		\caption*{(\#3) HD\,52961}\label{fig:ap_fitrt_chisq_hd52}
        	\end{subfigure}
        	\caption{Illustrations of the two-dimensional reduced chi-squared distribution for the grid of jet densities $n_j$ and temperature $T_j$. The white dot indicates the location of the best-fitting model. The contours represent the $1\,\sigma$, $2\,\sigma$, and $3\,\sigma$ intervals.}\label{fig:ap_fitrt_chisq_1}
        \end{figure*}

\begin{figure*}	
        	\begin{subfigure}[b]{.5\textwidth}
        		\centering\large 
        		\includegraphics[width=1\linewidth]{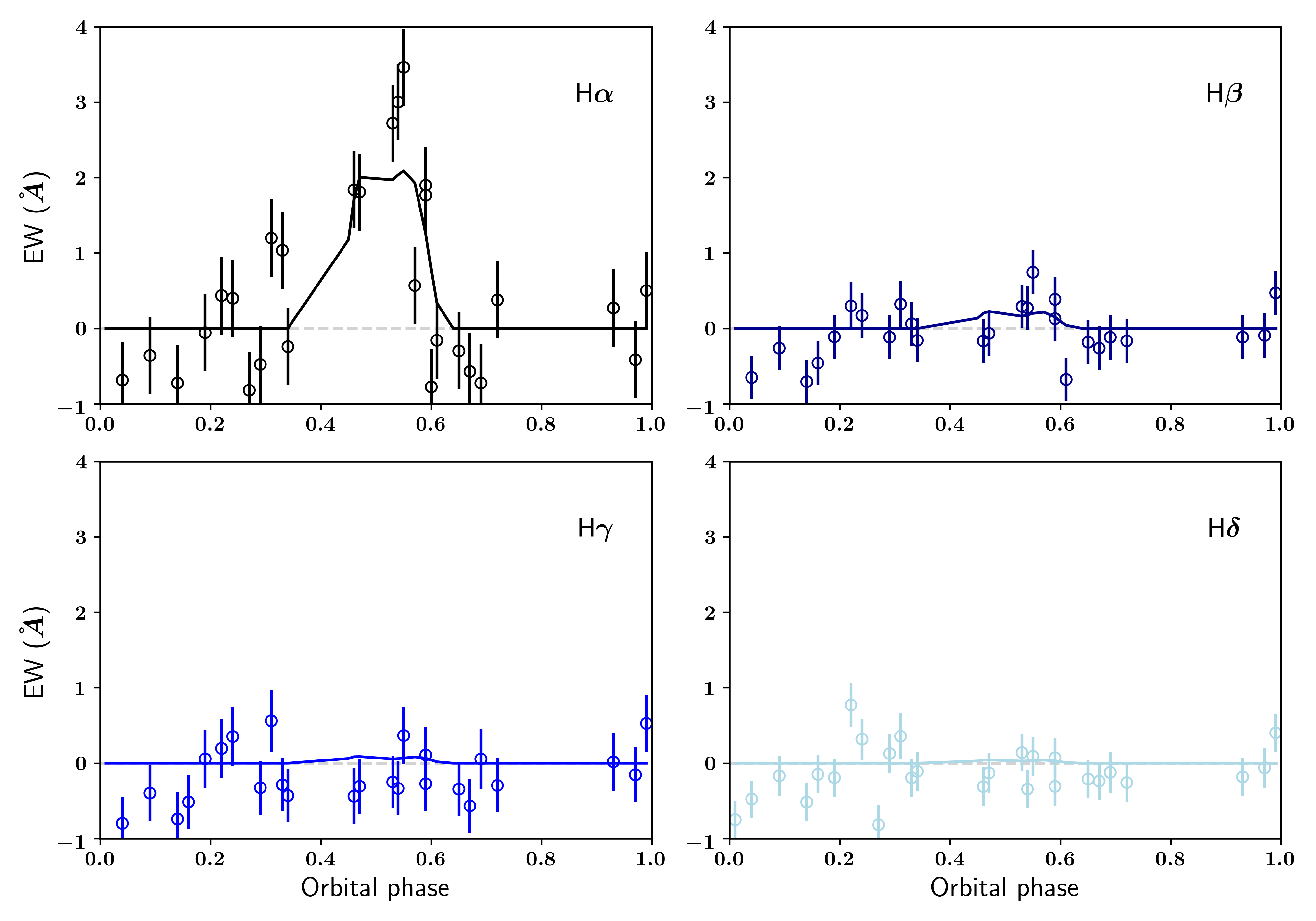}
        		\caption*{(\#1) AC\,Her}\label{fig:ap_fitrt_ew_ac}
        	\end{subfigure}%
        	\begin{subfigure}[b]{.5\textwidth}
        		\centering\large 
        		\includegraphics[width=1\linewidth]{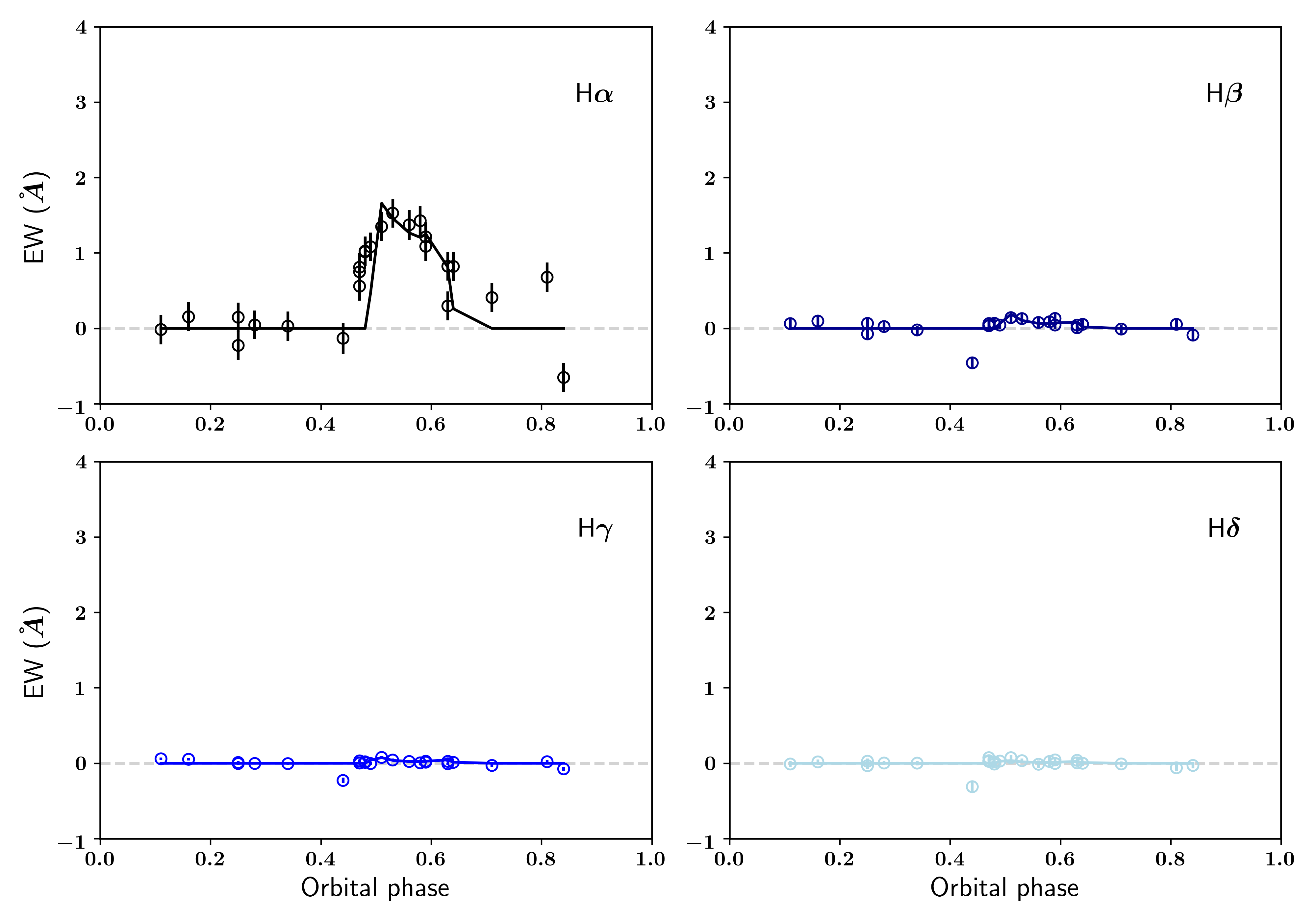}
        		\caption*{(\#2) HD\,213985}\label{fig:ap_fitrt_ew_hd21}
        	\end{subfigure}
       	\begin{subfigure}[b]{.5\textwidth}
        		\centering\large 
        		\includegraphics[width=1\linewidth]{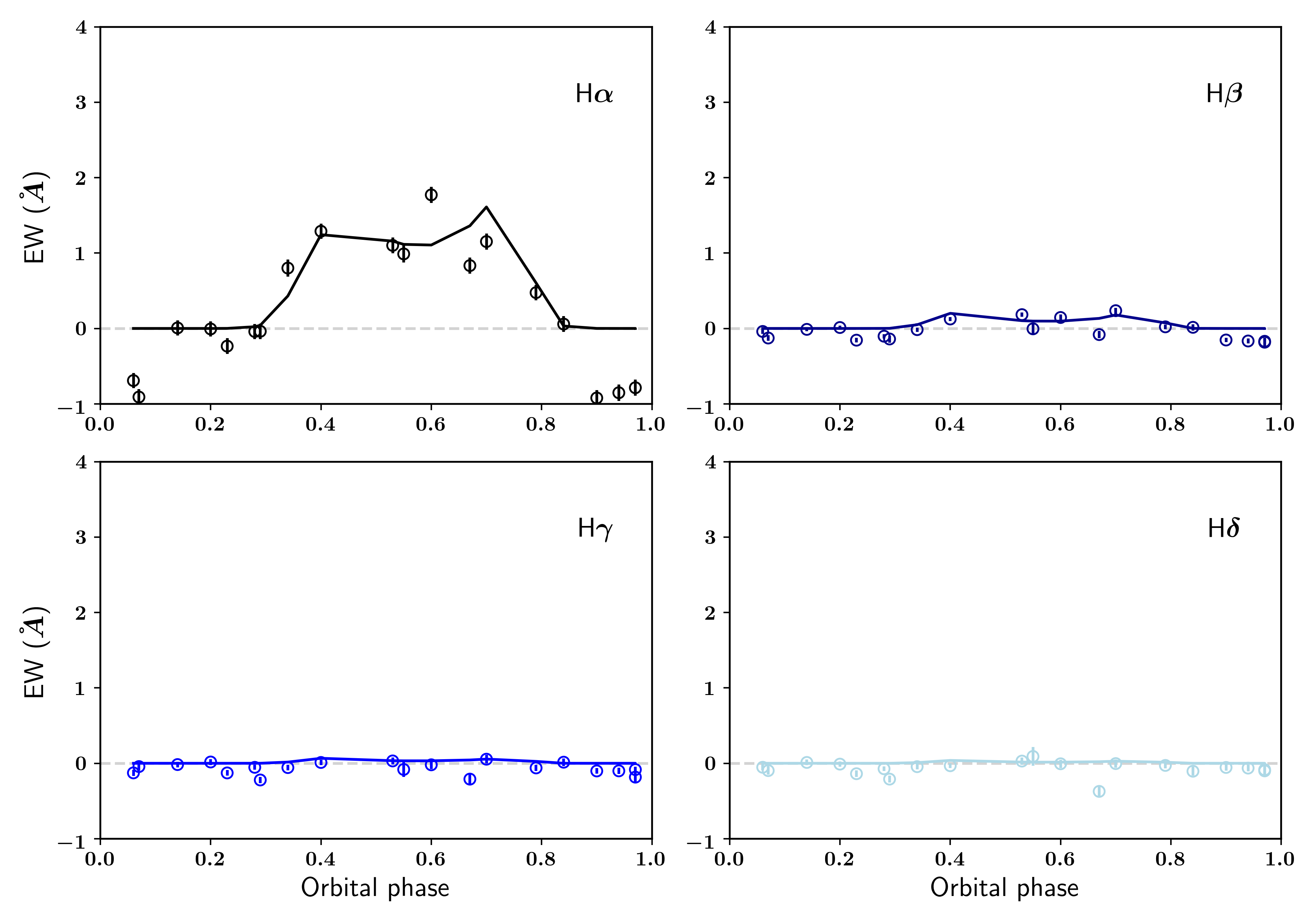}
        		\caption*{(\#5) IRAS19125+0343}\label{fig:ap_fitrt_ew_iras19125}
        	\end{subfigure}%
        	\begin{subfigure}[b]{.5\textwidth}
        		\centering\large 
        		\includegraphics[width=1\linewidth]{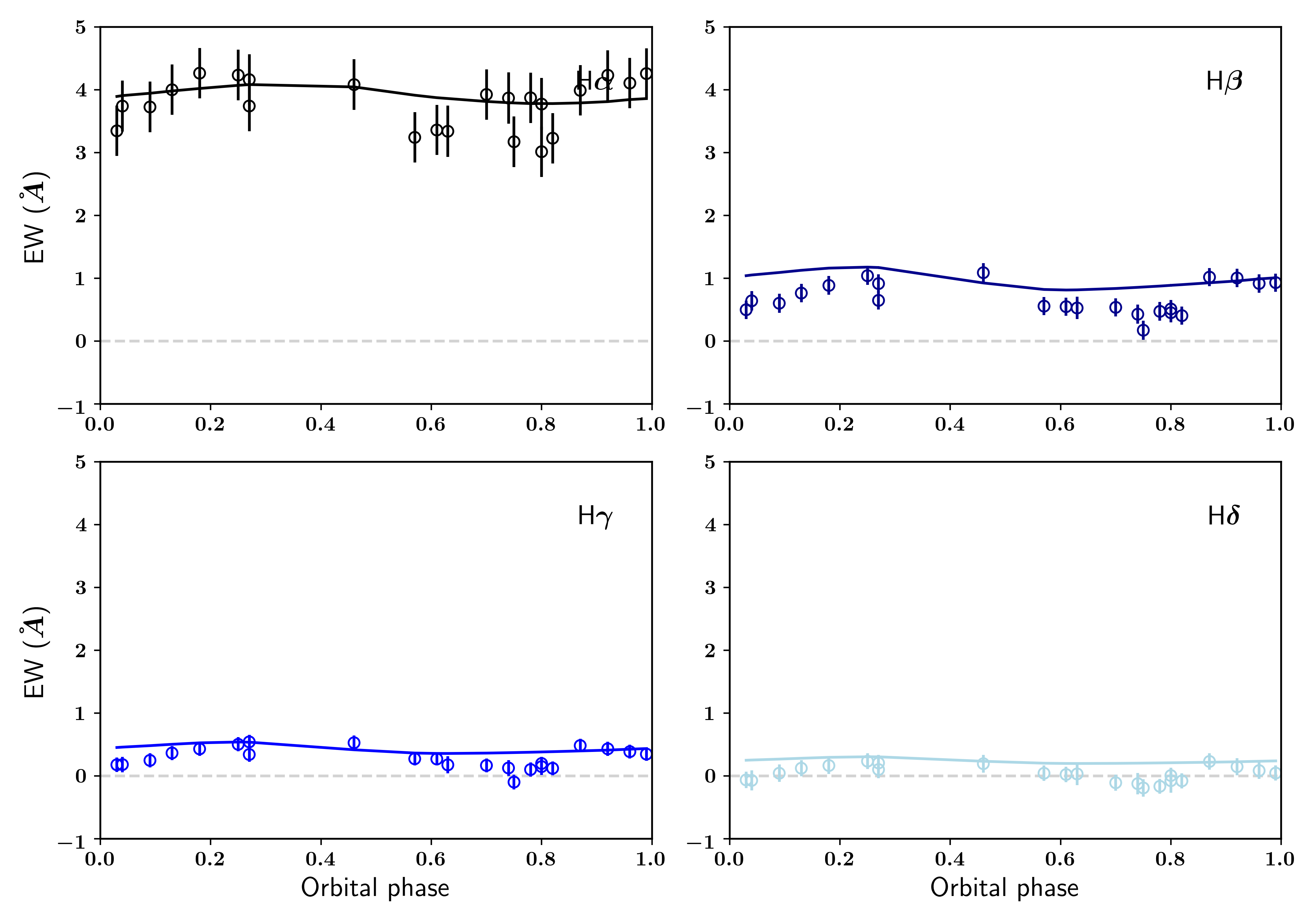}
        		\caption*{(\#7) SAO~173329}\label{fig:ap_fitrt_ew_sao}
        	\end{subfigure}
        
        	\caption{Illustration on the Balmer decrement fits in which the equivalent width of the absorption by the jet as measured on the data (\textit{dots}) is compared to the model estimates (\textit{full line}). The panels show the equivalent width in \halpha, \hbeta, \hgamma, and \hdelta.}\label{fig:ap_fitrt_89ac}
        \end{figure*}

        The results of the radiative transfer modelling are presented in Table~\ref{tab:rtresults}. Some examples of the 2D $\chi^2_\nu$\,distributions for jet densities and temperatures are presented in Fig.~\ref{fig:ap_fitrt_chisq_1}. These $\chi^2_\nu$\ values were determined using the equivalenth width determinations of the four Balmer lines and some of these equivalent width fits are shown in Fig.~\ref{fig:ap_fitrt_89ac}. Each panel in these figures shows the fit to the observed equivalent widths for the Balmer lines. In these plots, a positive equivalent width corresponds to a net absorption and a negative equivalent width corresponds to a net emission.  It can be seen that the equivalent width, and thus the strength of the jet absorption feature, is strongest for \halpha\,and weakens for each subsequent line in the Balmer series. This is caused by the higher number of hydrogen atoms in the $n=3$ energy state (which results in the $3\rightarrow2$ \halpha\,transition) than hydrogen atoms in higher energy states.

        In order to determine how much mass is lost by the jet, we need to know the jet geometry, velocity, and density structure. We calculate the jet mass-loss rates by combining the results from the spatio-kinematic modelling and the radiative transfer modelling. These estimates are presented in Table~\ref{tab:mrates}. We tabulate the upper and lower bonds which come from the propagation of the uncertainty of the model parameters and do not inlude systematics like corrections for non-LTE effects. Next, we estimate the mass accretion rate onto the companion for these objects. We assume a jet ejection efficiency of $\dot{M}_\text{jet, tot}/\dot{M}_\text{accr} = 0.4$\footnote{$\dot{M}_\text{jet,tot}$ refers to the mass loss rate from both jet lobes. The mass-loss rates calculated from the model are those for one lobe.}. The resulting accretion rates are also presented in Table~\ref{tab:mrates}.

        \begin{table*}
        	\begin{center}
        		\caption{Derived accretion and mass-transfer rates in the binary systems. The tabulated parameters are: the mass loss rate in both lobes of the jet $\dot{M}_\text{jet,tot}$ , the mass accretion rate feeding the circum-companion accretion disk $\dot{M}_\text{accr}$, the mass-transfer rate from the circumbinary disk $\dot{M}_\text{tr,CBD}$, and the mass-transfer rate from the post-AGB star $\dot{M}_\text{tr,pAGB}$. For each parameter, we show the average value and its upper and lower bound. }
        		\label{tab:mrates}
        		\scalebox{.85}{\begin{tabular}{ll c c c c c}\\
        				\hline \hline
        				\#  &  Object    &  & $\dot{M}_\text{jet,tot}$ &  $\dot{M}_\text{accr}$ &  $\dot{M}_\text{tr,CBD}$ & $\dot{M}_\text{tr,pAGB}$  \\
        				&  &   & $\myr$ &  $\myr$ & $\myr$ & $\myr$  \\
        				\hline
        				&  & Upper                & $ 1\times10^{-5}$   & $3\times10^{-5}$   & $1.2\times10^{-6}$   &  $4\times10^{-8}$\\
        				1  &  AC\,Her   & Average & $ 2.6\times10^{-6}$ & $7\times10^{-6}$   & $2\times10^{-7}$   &  $8\times10^{-9}$\\
        				&  & Lower                & $ 1\times10^{-6}$   & $3\times10^{-6}$   & $1.4\times10^{-8}$   & $8\times10^{-10}$\\
        				\hline
        				&  & Upper                & $ 6\times10^{-7}$   & $1.4\times10^{-6}$   & $1.0\times10^{-6}$   &  $1.0\times10^{-6}$\\
        				2  &  HD\,213985 & Average & $ 1\times10^{-7}$  & $3\times10^{-7}$   & $2\times10^{-7}$   &  $1.9\times10^{-7}$\\
        				&              & Lower   & $ 4\times10^{-9}$    & $9\times10^{-8}$   & $1.4\times10^{-8}$   & $1.9\times10^{-8}$\\
        				\hline
        				&  & Upper                & $ 1.2\times10^{-6}$   & $3\times10^{-6}$   & $1.6\times10^{-6}$   &  $1.5\times10^{-6}$\\
        				3  &  HD\,52961 & Average & $ 2.2\times10^{-7}$ & $6\times10^{-7}$   & $3\times10^{-7}$   &  $3\times10^{-7}$\\
        				&  & Lower                & $ 8\times10^{-8}$   & $2\times10^{-7}$   & $1.9\times10^{-8}$   & $3\times10^{-8}$\\
        				\hline
        				&  & Upper                     & $ 6\times10^{-3}$   & $1.3\times10^{-2}$   & $1.5\times10^{-6}$   &  $2\times10^{-9}$\\
        				4  &  IRAS06165+3158 & Average & $ 3.4\times10^{-4}$ & $9\times10^{-4}$   & $3\times10^{-7}$   &  $4\times10^{-10}$\\
        				&  & Lower                     & $ 2.2\times10^{-5}$ & $6\times10^{-5}$   & $1.8\times10^{-8}$   & $4\times10^{-11}$\\
        				\hline 
        				&  & Upper                     & $ 2.6\times10^{-6}$ & $7\times10^{-6}$   & $3\times10^{-6}$   &  $8\times10^{-7}$\\
        				5  &  IRAS19125+0343 & Average & $ 3.6\times10^{-7}$& $9\times10^{-7}$   & $6\times10^{-7}$   &  $1.5\times10^{-7}$\\
        				&  & Lower                     & $ 1.4\times10^{-7}$   & $3\times10^{-7}$   & $3\times10^{-8}$   & $1.5\times10^{-8}$\\  
        				\hline
        				&  & Upper                   & $ 1.6\times10^{-5}$     & $4\times10^{-5}$   & $2\times10^{-6}$   &  $2\times10^{-8}$\\
        				7  &  SAO~173329   & Average & $ 3.8\times10^{-6}$   & $1.0\times10^{-5}$   & $4\times10^{-7}$   &  $4\times10^{-9}$\\
        				&  & Lower                   & $ 1.6\times10^{-6}$     & $4\times10^{-6}$   & $2\times10^{-8}$   & $4\times10^{-10}$\\
        				\hline   
        				&  & Upper               & $ 6\times10^{-6}$         & $1.5\times10^{-5}$   & $5\times10^{-7}$   &  $4\times10^{-8}$\\
        				9  &  U\,Mon  & Average & $ 1.4\times10^{-6}$         & $3\times10^{-6}$   & $9\times10^{-8}$   &  $8\times10^{-9}$\\
        				&  & Lower               & $ 4\times10^{-7}$         & $1.2\times10^{-6}$   & $6\times10^{-9}$   & $8\times10^{-10}$\\
        				\hline
        		\end{tabular}}
        	\end{center}
        \end{table*}

\section{Discussion}\label{sec:discussion}
    As can be seen in the fits of the dynamic spectra in Fig.~\ref{fig:dynspec1} and the fits of the Balmer line equivalent widths in Fig.~\ref{fig:ap_fitrt_89ac}, we are able to fit the observed line profile variability well.  The fits provides us with crucial information about the jet (such as the jet structure), the binary system (such as the inclination angles), and the post-AGB star (such as the stellar radius). Our models do not include the pulsationally induced variability of the spectra, nor the potential cycle-to-cycle variability. 
    
    In this section, we discuss the overall structure of the jets in our systems. We also look into the mass-transfer history of these binary systems by comparing the estimated mass-accretion rates onto the companion with possible mass-transfer rates from the circumbinary disk and the post-AGB star to the companion. For the last part of our discussion, we explore correlations between the jet and properties of the circumbinary disk.


    \subsection{The jet morphology}\label{ssec:jetmorpology}
    
        We fitted three distinct jet configurations in the spatio-kinematic modelling of the jet, which represent three different jet-launching mechanisms: the stellar jet, the X-wind, and the disk wind configurations. Our fitting results marginally point to either a stellar jet or X-wind configuration, with four objects better fitting the former and three objects by the latter configuration. However, we confirm the conclusions of Paper~II, that we cannot robustly differentiate between these three launching models as the minimal $\chi^2$ values are similar. In Table~\ref{tab:ap_skmresults1} we give the model with the lowest $\chi^2$.

        The jet morphologies in our systems show, however, several key characteristics: they are all wide ($>30\degr$), with a slow and dense outflow component along the sides of the jet cone ($1-80\,$\kms) and a faster inner jet component ($>150\,$\kms). The innermost regions of the jets which favour a stellar jet configuration, resemble the jet cavity described in the X-wind and disk wind configurations. The cavity is a common feature in all our solutions but the density and velocity gradient between the symmetry axis and the cone's sides differs.  This can be seen in Figs.~\ref{fig:dyngeom} and~\ref{fig:dyngeom_updated}.

        The overall morphologies of the jets in our sample are very similar to the conical outflow morphology found by \cite{romanova09} in their 3D MHD simulations of jets launched from young stars. These conical jets are similar in nature to the X-wind, as both are launched from the inner disk regions. The main differences are that the conical outflow can be launched by stars with any rotation rate, whilst the X-wind is created by stars with high rotation \citep{shu94}.
        Moreover, the matter in the X-wind can be ejected in many directions, whereas for the conical outflow, the bulk of matter is ejected along a dense conical wind with a half-opening angle of $\sim30\degr$. \cite{romanova09} also showed that this conical outflow has a fast, low density jet component, which is located inside the dense conical outflow. 
        
       However, several sample stars have half-opening angles that are much larger than the half-opening angle of $30\degr$ (see Figs.~\ref{fig:dyngeom} and~\ref{fig:dyngeom_updated}). Thus, it is difficult to uniquely constrain the exact jet-launching mechanism responsible for the creation of jets found in our post-AGB binary systems.

    \subsection{Mass transfer in the binary system}\label{ssec:masstransfer}
    
        \begin{figure}[!htb]
        	\centering
        	\includegraphics[width=1\linewidth]{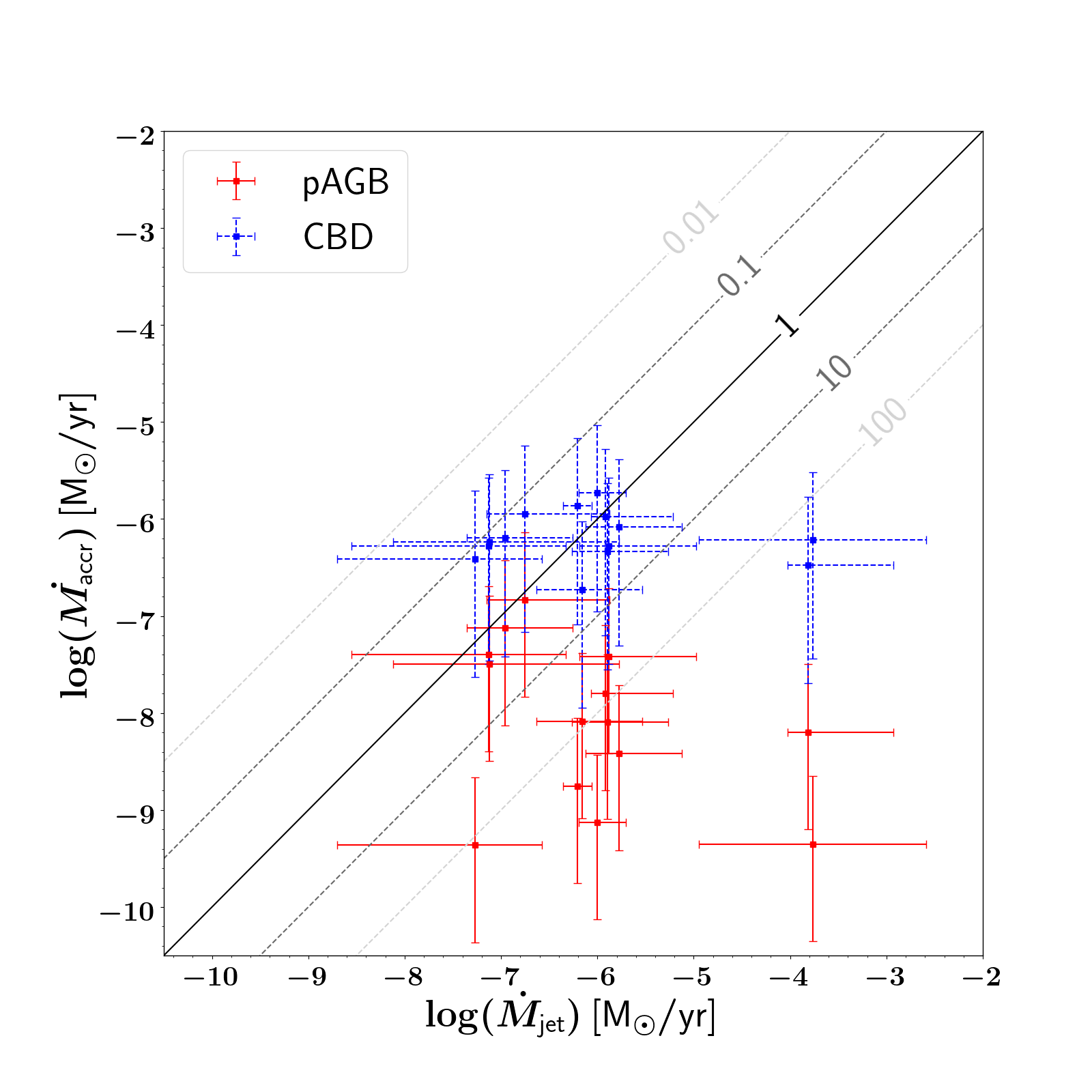}
        	\caption{Estimated mass-transfer rates onto the companion from the post-AGB star (solid red) and the circumbinary disk (dashed blue) vs. the jet mass-loss rates derived from the model for the objects in our sample (excluding RV\,Tau and SU\,Gem). The values are plotted on a logarithmic scale. The diagonal lines indicate the jet ejection efficiency ($\dot{M}_\text{jet}/\dot{M}_\text{accr}$). For instance a value of 0.1 indicates an ejection efficiency of $10\%$.}\label{fig:mtransfer}
        \end{figure}
        
        In Paper~I and ~II, we explored the link between the mass-loss rates in the jets and the origin of the  gas onto the disk around the companion. Mass that accretes onto the circum-companion disk must either flow from the post-AGB star via the inner Lagrangian point, or fall in from the circumbinary disk. The conclusion was that the circumbinary disk is likely the main source feeding the circum-companion accretion disk (see Paper~I and II).
        
        In this work, we have extended the analysis to all 14 objects in our sample for which we have an estimated mass accretion rate. We first estimate the mass loss rate of the post-AGB star and the accretion rate from the circumbinary disk to the central binary, for each object. We next compare these rates with the jet mass-loss rates that were determined from the modelling. 
        
        We follow a similar method to that described in Paper~I and II. For the mass transfer from the post-AGB star to the companion, we assume that mass transfer proceeds either via wind Roche lobe overflow \citep[WRLOF, ][]{mohamed07, abate13} or Bondi-Hoyle-Lyttleton (BHL) accretion \citep{hoyle39, bondi44, edgar04}. We rule out RLOF, because these systems do not show any ellipsoidal variations, which is expected if the post-AGB star fills its Roche lobe and undergoes RLOF \citep{kiss07, wislon76}. 
        
        The mass-transfer efficiency ($\dot{M}_\text{accr}/\dot{M}_\text{pAGB}$) for these two mass-transfer mechanisms is found to range from $1\,\%$ to $50\,\%$ \citep{devalborro09, abate13} with the higher values coming from WRLOF.  $\dot{M}_\text{accr}$ is the accretion rate feeding the accretion disk, and $\dot{M}_\text{pAGB}$ the current mass loss rate of the post-AGB primary.   
        
         We follow the prescription by \cite{schroder05} for the mass loss of the post-AGB stellar wind:
        \begin{multline}
            \dot{M}_\text{pAGB} = 8\times10^{-14}M_\odot/\text{yr} 
            \left( \frac{M_1}{M_\odot}\right)^{-1}
            \left( \frac{L_1}{L_\odot}\right)
            \left( \frac{R_1}{R_\odot}\right)
            \left( \frac{T_\text{eff}}{4000\,\text{K}}\right)^{3.5}\\
            \times\left( 1+\frac{g_\odot}{4300\,g_1}\right),
        \end{multline}
        
        \noindent with $M_1$, $L_1$, $T_\text{eff}$, and $g_1$ being the mass, luminosity, radius, effective temperature, and surface gravity of the post-AGB star, respectively.
        
        For the mass transfer rate from the circumbinary disk, we assume that $50\,\%$ of the re-accreted mass from the circumbinary disk is accreted and shared between both components equally:
        
        \begin{equation}
            \dot{M}_\text{tr, CBD} = \frac{1}{2}\frac{M_\text{0, CBD}}{t_0},
        \end{equation}
        where $M_\text{0, CBD}$ is the initial mass of the circumbinary disk and $t_0$ is the viscous time. The circumbinary disk mass is set to range between $6\times10^{-4}$ and $5\times10^{-2}M_\odot$, which are typical disk masses for post-AGB binary systems \citep{gielen07, bujarrabal13a, bujarrabal18, hillen17, kluska18}. We follow the prescription by \cite{rafikov16b} for the viscous time of the disk, which they define as
        \begin{equation}
            t_0 = \frac{4}{3}\frac{\mu}{k_B}\frac{a_b}{\alpha}\left[ \frac{4\,\pi\,\sigma\,(G M_{\rm tot})^2}{\zeta L_1} \right]^{1/4} \left( \frac{\eta}{I_L}\right)^2,
        \end{equation}
        where $\mu$ is the mean molecular weight, set to twice the proton mass, $k_B$ is the Boltzmann constant, $a_b$  is the binary separation, $\alpha=0.01$ is the viscosity parameter, $\sigma$ is the Stefan-Boltzmann constant, $G$ is the gravitational constant, $M_{\rm tot}$ is the total mass of the binary, $\zeta=0.1$ is a constant factor that accounts for the starlight that is intercepted by the disk surface at a grazing incident angle, $\eta=2$ is the ratio of angular momentum of the disk compared to that of the central binary, and $I_L=1$ accounts for the spatial distribution of the angular momentum in the disk. For our systems, $t_0$ ranges from 8000 to 55000 years with a typical value of 20000 years. We refer to \cite{oomen19} for a more detailed description of the time evolution of the expected accretion rate from a circumbinary disc.
    
        
        
        The estimated mass-accretion rates from the post-AGB star and the circumbinary disk to the companion are presented in Table~\ref{tab:mrates}. We compare these values with the jet mass-loss rates that are derived from the model in Fig.~\ref{fig:mtransfer}. The jet mass-loss rates range from $10^{-8}\,\myr$ to $10^{-4}\,\myr$. The diagonal lines indicate the jet ejection efficiency $\dot{M}_\text{jet}/\dot{M}_\text{accr}$. A value of 0.1 corresponds to an ejection efficiency of $10\%$. Ejection efficiencies larger than 1 ($100\%$) are not stable and indicate that either the mass accretion estimate is too low or the jet mass loss estimate is too high (or both). We show the lines of 10 and 100 times to guide the eye on the impact of these high values on the position of the objects.  We find that the mass-accretion rates from the post-AGB star are in general too low to account for the estimated jet mass-loss rates. However, mass transfer from the circumbinary disk is possible for all except two cases (including error bars). This strengthens the argument that was put forward in Paper~I and II, namely that re-accretion from the circumbinary disk is likely the primary source that feeds the accretion disk around the companion, from which the jet is launched.

        The lifetime of the circumbinary disk is mainly governed by the total disk mass and the mass loss rate. \cite{oomen19} found that the circumbinary disks must have a high disk mass of $\sim 10^{-2}\,\mdot$ in order to induce the observed depletion pattern of some post-AGB stars. The mass-accretion rates estimated in this work and papers Paper~I and II range between $10^{-7}\,\myr$ and $10^{-5}\,\myr$ (excluding IRAS06165+3158 and TW\,Cam). The lifetime of these disks can be estimated by $t_d = M_d / \dot{M}_a$. This results in disk lifetimes between $1,000$ and $100,000$ years, which is similar to the time a star spends between the AGB and the WD stage, when all nuclear burning stops.

         In two objects, IRAS06165+3158 and TW\,Cam, we estimate very high jet mass-loss rates, of the order of $\sim10^{-4}\,\myr$.
        The estimated mass-transfer rates from the circumbinary disk and the post-AGB stars are three and five orders of magnitudes smaller respectively (see Fig.\,\ref{fig:mtransfer}).
        Moreover, if the circumbinary disk would supply this mass, the disk lifetime would only be ten years, which is too short and inconsistent with observations. 
        
        Thus, it is more likely that jet mass loss estimate for these two objects are too high. One factor that could cause these high accretion rates is the jet temperature description in our jet model. The model assumes the jet temperature to be constant. Since the jet is observed as an absorption feature in the Balmer lines, it can be assumed that the jet temperature found by the model is cooler than the effective temperature of the star. For TW\,Cam and IRAS06165+3158, the effective temperatures are low ($4800\,$K and $4250\,$K), compared to the other objects in our sample, which would result in even lower jet temperatures. Since the jet temperature and density are mainly negatively correlated, a jet temperature estimate that is too low results in a jet density estimation that is too high. 
        
        We assume that the LTE approximation is valid for the excitation of hydrogen in the jet material. A potential increase of the excitation by non-LTE effects, will directly result in a decrease of absolute densities of our jet-model needed to fit the equivalent widths of our data. This leads to a decrease of the mass-loss estimates. Non-LTE effects are, however, not constrained by the data. 
        
        In future work, we will include a more complete description of the jet temperature structure and evaluate the impact of potential non-LTE excitation levels to evaluate if these could reduce the high mass accretion rate estimates. Additionally, we will make use of existing state-of-the-art 3D radiative transfer codes to model the Balmer lines in these systems \citep{hennicker20}. The results from our spatio-kinematic and radiative-transfer model-fitting routines can be used as input for the code. By doing so, we can assess how well our models perform in deriving the jet and binary properties in these systems.
        

    \subsection{The link between the circumbinary disk and the jet}\label{ssec:link_cbd_jet}
    
        We have established that the circum-companion disk from which the jet is launched, originates mainly from accreted material from the circumbinary disk. This indicates that the mass loss in the jet is strongly constrained by the properties of the circumbinary disk as the jets are likely produced from disk-binary interactions. A more massive circumbinary disk with an inner radius close to the central binary should result in a higher re-accretion rate, and thus, a higher jet mass loss rate. We also expect that disks with a large inner radius, will be less prone to efficient accretion. Hence, we expect that systems that have a strong jet absorption feature in \halpha\, will also show evidence for massive circumbinary disk with small inner radii.
        
         
\begin{figure}[!htb]
    \centering
    \includegraphics[width=1\linewidth]{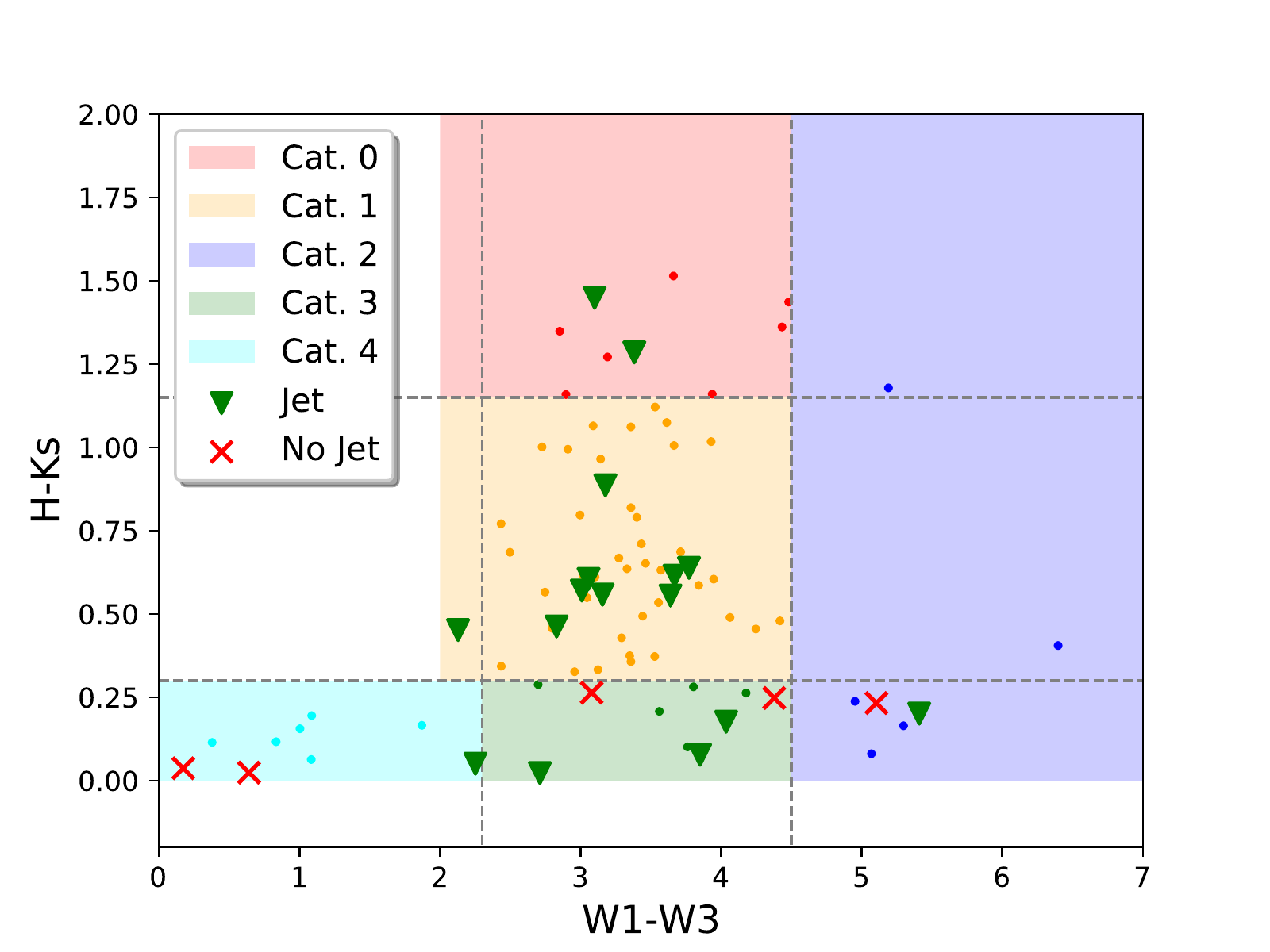}
    \caption{ Location of the objects studied in this paper in the infrared colour-colour diagram (the mid-IR WISE passbands W1-W3 versus H-Ks)  used in \citet{kluska22}. The areas covered by the different SED categories are coloured accordingly. The dashed lines are the limits used to determine the categories. 
    \label{fig:JetsVsSED}}
\end{figure}

        \cite{kluska22} have categorised the SEDs of the whole sample of Galactic post-AGB stars and link the observable colours to structural components of the disks. The SEDs are sensitive to the dust distribution of the circumstellar material, but are not as sensitive to the circumstellar gas component, although the gas pressure puffs up the disks to a significant scaleheight. We refer to appendix A of \cite{kluska22} for a graphical representation of the SEDs. These figures also include the SEDs of the objects discussed here.
        
        The majority of objects are surrounded by full disks \citep[category 0 and 1 in the classification of][]{kluska22}, which are dusty disks starting at the sublimation radius
        \citep{hillen16,kluska19} with a characteristic hot and puffed-up inner rim \cite[e.g.][]{hillen17,kluska18,corporaal21}. 
A smaller sub sample consists of objects that are called {\sl transition} disks (category 2 and 3), in which the inner dust radius is significantly larger than the dust sublimation radius. 
The SEDs of these objects have much less near-infrared excess than the full disks. The best studied such object is AC\,Her \citep{hillen15} where interferometric data is used to spatially resolve the large inner rim. Finally in a number of binaries, the dust excess is very weak (category 4) or not detected at all.  

\cite{kluska22} propose a disk evolutionary scenario in which the disks start as full disks and evolve to become transition disks till the dust is completely dispersed. We place the jet launching systems in this framework in Fig.~\ref{fig:JetsVsSED}. If the jets are fed by the circumbinary disk, we expect that objects with full disks have a jet, whereas binaries surrounded by an evolved transition disk, or without infrared excess will have a weaker jet or even no jet anymore.
        

        

We detect jets in all the eleven full disks objects in our sample. 
Indeed, all category 0 and 1 objects, this is with a rather high near-infrared colour (H-Ks>0.3), have a jet.
For targets with a weak infrared excess and low near-infrared colours (H-Ks<0.3), we detect a jet in only half of them (5/10).
All the five targets without a detected jet (see Section~2) have a low near-infrared colour, strengthening the hypothesis that the circumbinary disk feeds the jet.

\begin{figure}[!h]
    	\centering
    	\includegraphics[width=.5\textwidth]{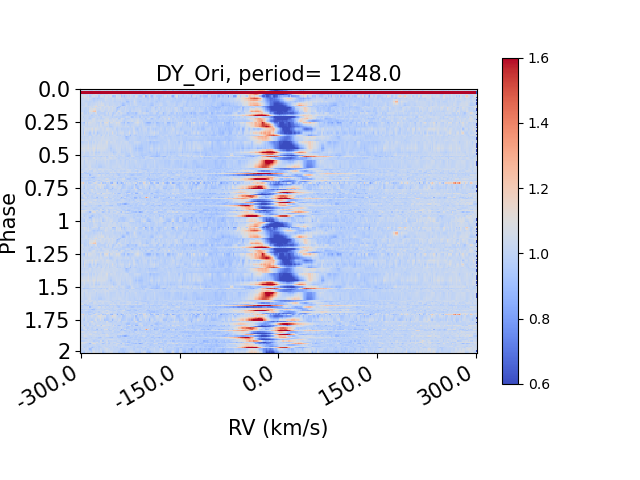}
    	\caption{Dynamic spectra of \halpha\, in the high-amplitude pulsator DY\,Ori, folded on its orbital period - two orbital periods are displayed. The spectra are normalised to the continuum. There is no clear jet feature detected in this object. }
    	\label{fig:dyori}
    \end{figure}

In the most obvious cases are HD\,137569 and BD+39$^\circ$4926 for which no, or only a very small, infrared excess is detected. They are located in the bottom right part of the colour-colour diagram and no jets are detected either.
However, we would have expected the same for HD\,46703 (SED category 4)  and EP\,Lyr, (SED category 3) stars that have a very small infrared excess (see Fig. A7 and A8 of \cite{kluska22} as IRAS06338+5333 and IRAS19163+2745 respectively), but instead a jet is detected in both systems. These objects still manage to accrete enough material to launch jets, despite their thinned out disks. On the other hand, DY\,Ori is an example of a transition disk with still a large infrared excess \citep[see IRAS06034+1354 = DY\,Ori in Fig. A6 of ][]{kluska22}, but in this object no clear jet is detected (see Fig.~\ref{fig:dyori}).
We conclude that, while a full disk seems to guarantee the presence of a jet, it is not yet clear what triggers or stops jets in objects with transition disks.

We can further speculate on the reason why some post-AGB stars have weaker jets, following the SED evolutionary scenario of \citet{kluska22}. They proposed that planets are the most likely explanation for the creation of cavities in the circumbinary disks. 
A giant planet, with a mass between 1 and 10\,M$_\mathrm{jupiter}$ can trap the dust in a pressure maximum outside its orbit and only let the gas through, which can be accreted onto the central stars. 
During such a process the planet will also accrete part of the gas and grow.
When growing, the planet may also open a gap in the gas and will, thus, stop the mass accretion onto the central stars.
Without accretion from the disk, the jet will stop. The presence or absence of  jets in evolved disks, could then depend on how efficiently the planet can stop the accretion process.

        
        

To summarise, there is a link between the circumbinary disk characteristics and the presence of jets. We detect jets in all systems with full disks (11/11) and only in half of the systems with weak or no near-infrared excess (5/10). 
The two disks with no infrared excess do not show any jet.
However, we have not found any clear explanation for why some of the transition disks do show a jet and others do not.
Spatially resolved data is needed to study the interaction between the inner regions of the circumbinary disk and the central binary were the jet is created by the accretion disk around the companion.



\section{Summary and conclusions}\label{sec:conclusions}
    
    In this work, we showed that we can obtain important information about the binary post-AGB systems by analysing the orbital phase dependent changes in the Balmer lines created by jets. We collected high-resolution optical spectra for a target sample of 9, jet-creating systems and applied the modelling techniques that were laid out in the Paper~I and II to these systems which are the spatio-kinematic and radiative transfer modelling. We obtained the spatio-kinematic structure of these jets, their mass-loss rates, and the mass-accretion rates in the circum-companion accretion disk. 

    Our spatio-kinematic model-fitting technique has proven to be successful in reproducing the observed variations in the \halpha\,lines for all objects in our sample. We find that the jets have wide half-opening angles ($>30\degr$). These wide opening angles are often found for jets launched by evolved low-mass stars, such as PNe and proto-PNe \citep{soker04,akashi13,akashi18}. In none of the systems we suspect a compact companion, as all jet velocities are typical for escape velocities of main sequence stars. Many jets in our sample show a significant tilt. This tilt could either be an indicator of jet precession, which is a common phenomenon for jets launched from binary systems, or jet deflection caused by the orbital motion of the binary system.
    
    We could not clearly distinguish between the three jet configurations: stellar jet, X-wind, and disk wind. The resulting jet structure fits turn to values very similar to one another, as they all have a dense and slow outer region with a faster inner jet region. This is also found in YSOs, where a slow extended wind is responsible for the bulk of the jet mass loss and a central faster wind is simultaneously present \citep{ferreira06, romanova09}. An important jet property that could be used to better constrain the jet-launching mechanism, and which is not yet included in our jet model, is jet rotation \citep{anderson03}. The jet rotation is highly dependent on the launching conditions of the jet \citep{ferreira06}. Thus, by incorporating a toroidal velocity component in our jet model, we might be able to better constrain the jet-launching mechanism that governs jets in post-AGB binary systems.  Including jet rotation and studying the impact it has on the observables in our model will be part of future work.
    
    We combined the results of the spatio-kinematic and radiative-transfer modelling to determine mass-loss rates in the jet. These jet mass-loss rates are high ($10^{-8}-10^{-5}\,\myr$), from which we can infer that the mass-accretion rates onto the companion star must be high as well.  By comparing mass transfer from the post-AGB star and re-accretion from the circumbinary disk with the mass-accretion rates in the disk, we find that the circumbinary disk must be the main source of mass that sustains the observed jet creation from the circum-companion disk.
    
    We found that our radiative transfer model predicts unrealistically high mass-loss rates especially for the jets in IRAS06165+3158 and TW\,Cam, which are the coolest objects in our sample. The jets are still seen in absorption, and in our model, this means that the jet must be cool as well. The mass-loss rates found by the model in these systems are on the order of $>10^{-4}\,\myr$ and cannot be explained by either disk re-accretion or binary mass transfer. These high estimates could be caused by the assumption of a constant jet temperature in the jet model. Our analyses was performed in LTE and a potential non-LTE excitation of hydrogen, will directly impact on our mass-loss estimates as well. In future work, we will include a more realistic jet temperature description in our model and will also use other more sophisticated 3D radiative transfer codes to model these systems as well possible deviations from LTE excitation patterns in the hydrogen atoms of the jet. By comparing the resulting line profiles with our existing model, we can assess the performance of our radiative transfer model.
    
    In this study, we showed how the derived mass-loss rates in the jet and several stellar and disk properties suggest that re-accretion from the circumbinary disk is most likely responsible for the observed jet mass loss. 
    The comparison of our results with characteristics of the circumbinary disks as given in \cite{kluska22} strengthen the hypothesis that the jets are fed by matter from the circumbinary disk. All systems with full disks show jets. For systems with transition disks (these are disks with an inner radius larger than the sublimation radius), only 50\% of them in our spectroscopic sample show jets pointing towards a mechanism stopping accretion from the disk to the central stars.  
    
   We aim to directly observe accretion streams from the inner disk rim to the central binary system by means of high-angular resolution imaging, which can give direct evidence of disk re-accretion in these post-AGB binary systems. Moreover, we plan to include self-consistent physical jet models instead of geometric models so as to test which physical models can account for the jets in post-AGB binaries. With the constrained velocities and densities, we conclude that the jets are an important ingredient in the structure of many post-AGB binaries and their impact on both the evolution and the shaping of circumstellar material is yet to be studied.


\begin{acknowledgements}
    This work was performed on the OzSTAR national facility at Swinburne University of Technology. OzSTAR is funded by Swinburne University of Technology and the National Collaborative Research Infrastructure Strategy (NCRIS). DK  acknowledges  the  support  of  the  Australian  Research Council  (ARC)  Discovery  Early  Career  Research  Award (DECRA) grant (DE190100813). DB and HVW acknowledge support from the Research Council of the KU Leuven under grant number C14/17/082. J.K. acknowledges support from FWO under the senior postdoctoral fellowship (1281121N). The observations presented in this study are obtained with the HERMES spectrograph on the Mercator Telescope, which is supported by the Research Foundation - Flanders (FWO), Belgium, the Research Council of KU Leuven, Belgium, the Fonds National de la Recherche Scientifique (F.R.S.-FNRS), Belgium, the Royal Observatory of Belgium, the Observatoire de Gen\`eve, Switzerland and the Th\"uringer Landessternwarte Tautenburg, Germany. 
\end{acknowledgements}


\bibliographystyle{aa}

\bibliography{allpapers.bib}


\end{document}